\documentclass[12pt]{article}
\usepackage{amsmath}
\usepackage{blkarray}
\usepackage[utf8]{inputenc}
\usepackage{csquotes}
\usepackage{hyperref}
\usepackage{comment}
\usepackage{titlesec}
\usepackage{verbatim}
\usepackage[linesnumbered,ruled,vlined]{algorithm2e}
\usepackage{graphicx}
\usepackage{amssymb}
\usepackage{latexsym}
\usepackage{amsfonts}
\usepackage[normalem]{ulem}
\usepackage{soul}
\usepackage{array}
\usepackage{amssymb}
\usepackage{extarrows}
\usepackage{subfig}
\usepackage{wrapfig}
\usepackage{txfonts}
\usepackage{wasysym}
\usepackage{enumitem}
\usepackage{adjustbox}
\usepackage{ragged2e}
\usepackage[svgnames,table]{xcolor}
\usepackage{tikz}
\usepackage{longtable}
\usepackage{changepage}
\usepackage{setspace}
\usepackage{hhline}
\usepackage{multicol}
\usepackage{tabto}
\usepackage{float}
\usepackage{multirow}
\usepackage{makecell}
\usepackage{fancyhdr}
\usepackage[title,titletoc,page]{appendix}
\usepackage{authblk}
\usepackage{flowchart}
\usepackage{amssymb}
\usepackage{lipsum}
\usepackage{mathrsfs}
\usepackage[style=apa,backend=biber,maxcitenames=2]{biblatex}
\usepackage[margin=0.75in]{geometry}
\usepackage[T1]{fontenc}
\usepackage{bm}

\usepackage{blindtext}
\makeatletter
\renewcommand\maketitle
  {\centering
   {\Large\bfseries\@title}%
   \medskip\par\centering
   {\large\bfseries\@author}%
  }
\makeatother

\DeclareNameAlias{sortname}{family-given}
\DeclareFieldFormat[article,periodical]{volume}{\mkbibbold{#1}}
\DeclareMathOperator*{\argmaxA}{arg\,max} 
\DeclareMathOperator*{\blockdiagA}{blockdiag} 

\AtEveryBibitem{%
\clearfield{number}}
\renewbibmacro{in:}{}

\newcommand*{\permcomb}[4][0mu]{{{}^{#3}\mkern#1#2_{#4}}}
\newcommand*{\comb}[1][-1mu]{\permcomb[#1]{C}}

\newtheorem{theorem}{Theorem}
\newtheorem{proof}{Proof}
\newtheorem{coll}{Corollary}

\TabPositions{0.5in,1.0in,1.5in,2.0in,2.5in,3.0in,3.5in,4.0in,4.5in,5.0in,5.5in,6.0in}
\setlistdepth{9}
\renewlist{enumerate}{enumerate}{9}
		\setlist[enumerate,1]{label=\arabic*)}
		\setlist[enumerate,2]{label=\alph*)}
		\setlist[enumerate,3]{label=(\roman*)}
		\setlist[enumerate,4]{label=(\arabic*)}
		\setlist[enumerate,5]{label=(\Alph*)}
		\setlist[enumerate,6]{label=(\Roman*)}
		\setlist[enumerate,7]{label=\arabic*}
		\setlist[enumerate,8]{label=\alph*}
		\setlist[enumerate,9]{label=\roman*}

\renewlist{itemize}{itemize}{9}
		\setlist[itemize]{label=$\cdot$}
		\setlist[itemize,1]{label=\textbullet}
		\setlist[itemize,2]{label=$\circ$}
		\setlist[itemize,3]{label=$\ast$}
		\setlist[itemize,4]{label=$\dagger$}
		\setlist[itemize,5]{label=$\triangleright$}
		\setlist[itemize,6]{label=$\bigstar$}
		\setlist[itemize,7]{label=$\blacklozenge$}
		\setlist[itemize,8]{label=$\prime$}

\bf\title{Model-robust Bayesian design through Generalised Additive Models for monitoring submerged shoals}

\author{De Silva, D.$^{*1}$}
\author{Fisher, R.$^2$}
\author{Radford, B.$^2$}
\author{Thompson, H.$^{1,3}$}
\author{McGree, J. M.$^{1,3}$}
\affil{$^*$Corresponding author (email: dilishiya.desilva@qut.edu.au)}
\affil{$^1$School of Mathematical Sciences, Queensland University of Technology, Brisbane, QLD 4000, Australia}
\affil{$^2$Australian Institute of Marine Science, UWA Oceans Institute, Crawley, WA 6009, Australia}
\affil{$^3$Centre for Data Science, Queensland University of Technology, Brisbane, QLD 4000, Australia}

\begin{document}

\maketitle

\begin{abstract}

Optimal sampling strategies are critical for surveys of deeper coral reef and shoal systems, due to the significant cost of accessing and field sampling these remote and poorly understood ecosystems.
Additionally, well-established standard diver-based sampling techniques used in shallow reef systems cannot be deployed because of water depth. Here we develop a Bayesian design strategy to optimise sampling for a shoal deep reef system using three years of pilot data. 
Bayesian designs are generally found by maximising the expectation of a utility function with respect to the joint distribution of the parameters and the response conditional on an assumed statistical model. 
Unfortunately, specifying such a model {\it a priori} is difficult as knowledge of the data generating process is typically incomplete.  To address this, we present an approach to find Bayesian designs that are robust to unknown model uncertainty.  
This is achieved through couching the specified model within a Generalised Additive Modelling framework and formulating prior information that allows the additive component to capture discrepancies between what is assumed and the underlying data generating process. 
The motivation for this is to enable Bayesian designs to be found under epistemic model uncertainty; a highly desirable property of Bayesian designs.  
Our approach is demonstrated initially on an exemplar design problem where a theoretic result is derived and used to explore the properties of optimal designs. 
We then apply our approach to design future monitoring of sub-merged shoals off the north-west coast of Australia with the aim of significantly improving on current monitoring designs.
\end{abstract}

\justifying

\justifying

\section{Introduction} \label{section:introduction}

Understanding ecosystem states and trajectories requires a survey design that provides precise information from an efficient use of resources for monitoring. 
This is true for  many environmental domains including marine seascapes  \parencite{swartzman1997analysis,beare2002spatio,taylor2003spatial,winter2007variations,murase2009application}, plant biomes   \parencite{yee1991generalized} and atmospheric systems  \parencite{aldrin2005generalised,pearce2011quantifying}. 
The challenge of such data collection is to ensure the design remains efficient for the goal of the subsequent analysis despite potentially incomplete knowledge of the data generating process. 
Approaches to find designs that are robust to model uncertainty have been proposed in the literature  \parencite{cook1982model,sacks1984some,li2000model,yue2002model,kristoffersen2020model,selvaratnam2021model,xu2021robust} but rely on the true model being specified {\it a priori} in one form or another. 
In addition to being impracticable, the resulting design can yield misleading information if the assumed model is misspecified  \parencite{chang199628}.

In this paper, we consider Generalised Additive Models (GAMs)  \parencite{hastie1987generalized,hastie1990generalized} as a basis for specifying prior information about the model and the range of potential discrepancies between a prediction and the mean response that may be observed. 
GAMs are extensions of Generalised Linear Models (GLMs) that include smooth functions of predictor variables to provide flexibility in describing the relationship between the mean response and the covariates. 
Here, we propose to exploit this flexibility to provide Bayesian designs that are robust to unknown model uncertainty.

Throughout this paper, we consider finding designs within a Bayesian framework. 
There have been a number of approaches proposed to form model robust designs in such settings.
Most commonly, model robust Bayesian designs are formed through averaging a utility function over a finite set of plausible models \parencite[e.g.][]{zhou2003bayesian,bornkamp2011response,drovandi2014sequential}, an idea proposed by  \textcite{lauter1974experimental,lauter1976optimal}. 
This approach assumes that one of the models within the set will appropriately describe the data i.e.\ the M-closed perspective. 
While offering robustness to the form of the model, the approach still relies on being able to appropriately specify the data generating process {\it a priori}.  
The challenge we address here is to find designs within the M-open perspective.  
Research in this space appears to stem from  \textcite{welch1983mean} who included an additive term within a linear model to account for discrepancies between the assumed model and the true underlying mean response. 
 \textcite{sacks1984some} extended this idea to find designs within a class of possible linear models where the class was not finite dimensional.  
This led to the construction of designs for estimating characteristics of a non-parametric model providing robustness to unknown model uncertainty in linear settings.  

Design under additive versions of spline models has been considered more recently  \parencite{biedermann2009optimal,biedermann2011optimal,dette2008optimal, dette2011optimal}. 
If the number of knots is assumed to be known, then the design problem reduces to finding designs for segmented univariate polynomial regression models  \parencite{heiligers1998optimal,heiligers1999experimental,woods2003designing}.
This was extended to include the precise estimation of the number of knots where it was found that D-optimal designs were not necessarily robust to prior specification of the number of knots  \parencite{dette2008optimal}. 
Most recently,  \textcite{wang2020optimal} considered D-optimal designs while setting a prior on the curvature as a weighted smoothness penalty, and found that optimal design points appeared where there were large curvature values and on the design boundary of the design space. 
Of interest, it was found that design points appear to become more spread out as assumed model discrepancy increased.  
Intuitively, the more spread out design points are, the more robust the design would seem to be to model misspecification.  
As these recent efforts focus on linear additive models and Fisher information, a generalised approach for forming model-robust Bayesian designs has not been proposed.  
Through this paper, we aim to address this by providing an approach to find robust designs within a variety of settings.  


\subsection{Motivating example: Monitoring submerged shoals off the north-west coast of Australia} \label{subsection:motivating_example}

With coral reefs in decline worldwide  \parencite{hughes2018spatial}, there is an increasing interest in the role of deep  coral ecosystems and  submerged shoals  \parencite{bridge2013call}. 
This interest is motivated by an emerging understanding that they represent biodiversity hotspots  \parencite{moore2017submerged}, need an updated conservation status  \parencite{moore2016improving}, and may play a role in sustaining shallower reefs through, for example, larvae replenishment for fish and coral species. 
Deeper reefs in the remote parts of Australia are thought to be more resilient to decline as their isolation and offshore location protect them from some  anthropogenic and climatic disturbances. 
For example, their relative water depth compared to shallow reefs means they are less likely to experience coral bleaching due to reduced exposure to higher temperatures and ambient light. 
This has led to a call for greater ``information on the prevalence and the ecological and economic importance" of these deeper reefs  \parencite{bridge2013call}.

\begin{figure}[H]
    \centering
    \includegraphics[scale=0.75]{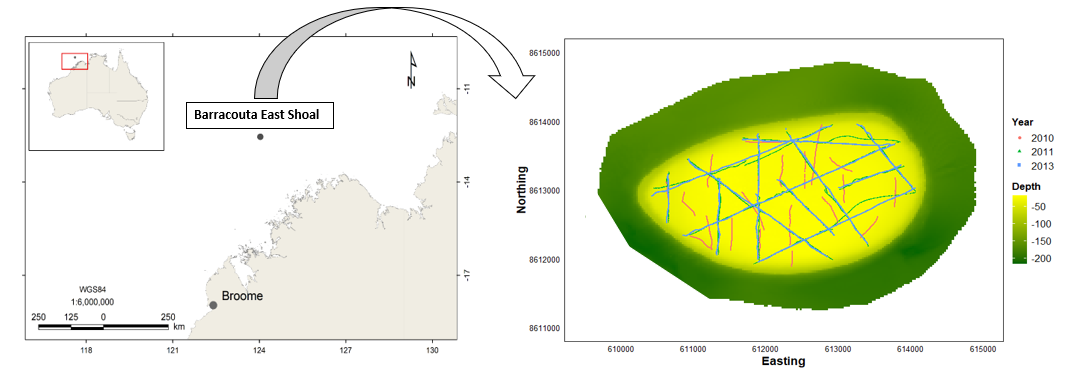}
    \caption{Map of the bio-region and the transects used for data collection at Barracouta East shoal across three sampling years 2010, 2011 and 2013.}
    \label{figure:AIMS_Transect_design_BE}
\end{figure}

The Australian Institute of Marine Science (AIMS) has been conducting surveys of submerged shoals off the western coast of Australia since 2010. 
The sampling design used for much of this initial exploration was based on a whole shoal approach with large transects covering the full depth ranges (from $-18$ to $-60$ m) from one side of a shoal to the other. 
These transects were purposefully positioned to transect depth gradients that delineate habitat boundaries and is a design that is generally used for developing spatial habitat maps. 
The data collected from these transects includes hard coral cover, and a range of covariates (see Appendix \ref{appendix:covariatedescription}). 
An example of the sampling that has been undertaken is shown in Figure \ref{figure:AIMS_Transect_design_BE} which shows the transects used to collect data at the Barracouta East shoal in $2010, 2011$ and $2013$ (for more information see,  \textcite{heyward2017barracouta}). 

Interest now is moving from mapping these submerged shoals to monitoring them, and in particular monitoring coral cover within a shoal. 
However, accessing these shoals for monitoring is very expensive and logistically difficult. 
An effective way to characterise the state and dynamics of key habitats in these systems is critical to improve ecological understanding and management. 
Optimal sampling strategies are hampered by a lack of understanding of the spatial dynamics of these systems, and the fact that the well-established diver-based  sampling techniques used in shallow reef systems cannot be deployed at depth. 
We start by providing the first published statistical analysis of these transect data by developing an appropriate
Generalised Additive Mixed Model, (GAMM)  \parencite{lin1999inference} where GAMMs are further extensions of the GAM framework to allow variability (say) spatially and between subgroups to be captured  \parencite{wood2006low}. 
We then apply our proposed design methods to find transects for data collection on this shoal, and explore the robustness properties of these designs.

The remainder of the paper is set out as follows. 
In Section \ref{section:modelling}, we provide background on Bayesian modelling for Generalised Additive (Mixed) Models (GA(M)Ms).  
In Section \ref{section:bayesiandesign}, our methodology for finding Bayesian designs for GA(M)Ms is described. 
This methodology is then applied in Section \ref{section:examples} via an illustrative example where a theoretic result is derived and then used to explore the properties of designs for a special case of GAM.  
Following this, the developed methodology is applied to find transects for monitoring submerged shoals. 
The paper then concludes with a discussion of the main outcomes of this research and suggestions for future studies.


\section{Statistical modelling framework} \label{section:modelling}

\subsection{Generalised Additive (Mixed) Models} \label{subsection:gamm}

To define a GA(M)M, let $\bm{y}=(y_1,\ldots,y_n)^T$ be the observed data collected with covariate information $\bm{X}=[\bm{1},\bm{x}_1,\ldots,\bm{x}_{(p+q)}]$ where $\bm{x}_j = (x_{1j},\ldots,x_{nj})^T$, for $j = 1,\ldots,(p+q)$. 
Assume the $p+q$ covariates are comprised of $p$ covariates that potentially have a non-linear relationship with the mean response and $q$ covariates whose influence on the mean response is assumed to be linear.  
Also assume that there may be random effect $\bm{s}$ within the model which could account for different sources of variability. 
Then a GA(M)M with mean $\bm{\mu}=\mathop{\mathbb{E}} (\bm{y})$ and response variable $\bm{y} \sim \mbox{EF}(\bm{\eta},\psi)$  where $\mbox{EF}(\bm{\eta},\psi)$ denotes a distribution from the exponential family (e.g. Bernoulli, Binomial, Poisson, Normal and Exponential) with additive semi-parametric predictor $\bm{\eta}$ and scale parameter $\psi$, can be expressed as follows:
\begin{equation}\label{Eq:1}
    \begin{split}
        g(\bm{\mu}) = {}& \bm{\eta}, \\
        \bm{\eta} = {\beta_0 + \sum_{j=1}^{p} f_j(\bm{x}_{j};\beta_j,\bm{u}_j)} +
        \sum_{j=p+1}^{q}\beta_{j}\bm{x}_{j} + {}&
        \sum_{a=1}^{p+q-1} \sum_{b=a+1}^{p+q} f_{x_a,x_b}(\bm{x}_{a},\bm{x}_{b};\bm{v}_{a,b}) +  \bm{s},
    \end{split}
\end{equation}

\noindent where $g(.)$ is a known link function (e.g. logit, log and identity). Here $\beta_j, j=0,\dots,(p+q)$, $\bm{u}_j, j=1,\dots,p$ and $\bm{v}_{a,b}, a=1,\dots,(p+q+1), b=(a+1),\dots,(p+q)$ denote model parameters, $f_j$'s are non-parametric functions that are assumed to be smooth, and $f_{x_a,x_b}$'s are functions for evaluating two-way interactions.

There are numerous ways by which the $f_j$'s can be defined and estimated. 
In this work, we consider penalised splines introduced by  \textcite{o1986statistical} which can be expressed as follows: 
\begin{equation}\label{Eq:2}
    \begin{split}
    \sum_{j=1}^{p} f_j(\bm{x}_{j};\beta_j,\bm{u}_j) = {}& \beta_{1}\bm{x}_{1}+ \sum_{k=1}^{K_1} u_{1k}z_{1k}(\bm{x}_1) 
    + \ldots + \beta_{p}\bm{x}_{p}+ \sum_{k=1}^{K_p} u_{pk}z_{pk}(\bm{x}_p),  \\
        = {}& \sum_{j=1}^{p} \left[ \beta_{j}\bm{x}_{j}+ \sum_{k=1}^{K_j} u_{jk}z_{jk}(\bm{x}_j) \right],
    \end{split}
\end{equation}

\noindent where for $j=1,\dots,p$, $K_j$ is number of knots for $j^{th}$ covariate, $u$'s are wiggliness parameters and $z$'s are orthogonalised O’Sullivan spline bases (see  \textcite{wand2008semiparametric}). 
O’Sullivan penalised splines are similar to P-splines (see  \textcite{eilers1996flexible}), and are a generalisation of smoothing splines. 

In many situations, an additive predictor (as shown in Equation \eqref{Eq:2}) may not appropriately capture the variability in the response due to interactions between covariates. 
Thus, interactions can be included in the model, and this can be achieved via tensor product smoothers. 
Here, for constructing low-rank tensor product smoothers to use as components of the GA(M)M, we use the method introduced by  \textcite{wood2006low} (further details in  \textcite{wood2017generalized}).
This method can be used to construct interactions between any number of covariates, and here we consider up to two-way interactions between, say, covariates $x_a$ and ${x_b}$, denoted by $f_{x_a,x_b}$. 
The process starts by assuming that we have low rank bases available for representing smooth functions of covariates $x_a$ and $x_b$ defined as follows:
\begin{equation}\label{Eq:3}
    f_{x_a}(\bm{x}_a;\bm{v}_{a}) = \sum_{k_{a}=1}^{K_{a}} v_{a_{k_{a}}} w_{k_{a}}(x_a) \quad\text{and}\quad f_{x_b}(\bm{x}_b;\bm{v}_{b}) = \sum_{k_{b}=1}^{K_{b}} v_{b_{k_{b}}} w_{k_{b}}(x_b),
\end{equation}

\noindent where the $v$'s are parameters and the $w$'s are known basis functions. Now, $f_{x_a}$ needs to vary smoothly with $x_b$. To accomplish this, the smooth parameters, $v_{a_{k_{a}}}$, need to vary smoothly with $x_b$. This could be written as:
\begin{equation}\label{Eq:4}
    v_{a_{k_{a}}}(x_b) = \sum_{k_{b}=1}^{K_{b}} v_{b_{k_{a},k_{b}}} w_{k_{b}}(x_b),
\end{equation}

\noindent then $f_{x_a,x_b}(\bm{x}_a,\bm{x}_b;\bm{v}_{a,b})$ can be written as follows:
\begin{equation}\label{Eq:5}
    f_{x_a,x_b}(\bm{x}_a,\bm{x}_b;\bm{v}_{a,b}) = \sum_{k_{a}=1}^{K_{a}}\sum_{k_{b}=1}^{K_{b}} v_{b_{k_{a},k_{b}}} w_{k_{b}}(x_b) w_{k_{a}}(x_a).
\end{equation}

Given appropriate ordering of $v_{b_{k_{a},k_{b}}}$ into a vector $\bm{v}_{a,b}$, if $\otimes_r$ is the row-wise Kronecker product, then for any particular set of observations of the two covariates say $x_a$ and $x_b$, the relationship between the model matrix, $\bm{W}_{a,b}$ which evaluates the tensor product smooth at these observations, and the model matrices $\bm{W}_a$ and $\bm{W}_b$ that would evaluate the marginal smooths at the same observations can be written as follows: 
\begin{equation}\label{Eq:6}
    \bm{W}_{a,b}=\bm{W}_a\otimes_r\bm{W}_b. 
\end{equation}

\sloppy Throughout, it will be convenient to describe the model in Equation \eqref{Eq:1} in matrix form. For this, let $\beta_0 + \sum_{j=1}^{p} f_j(\bm{x}_{j};\beta_j,\bm{u}_j) + \sum_{j=p+1}^{q}\beta_{j} x_{j}$ be written as $\bm{X\beta} + \bm{Zu}$. Here, $\bm{Z}=\big[\bm{Z}(\bm{x}_1),\bm{Z}(\bm{x}_2),\dots,\bm{Z}(\bm{x}_p)\big]$ where $\bm{Z}(\bm{x}_j)=\big[\bm{Z}_1(\bm{x}_j),\bm{Z}_2(\bm{x}_j),\dots,\bm{Z}_{K_j}(\bm{x}_j)\big], j=1,\dots,p$ and  $\bm{Z}_k(\bm{x}_j)=\big(z_{jk}(x_{j1}),z_{jk}(x_{j2}),\dots,z_{jk}(x_{jn})\big)^T$, $k=1,\dots,K_j$. Then, let $\sum_{a=1}^{p+q-1} \sum_{b=a+1}^{p+q} f_{x_a,x_b}(\bm{x}_a,\bm{x}_b;\bm{v}_{a,b})$ be written as $\bm{Wv}$ where $\bm{W}=\big[\bm{W}_{1,2},\dots,\bm{W}_{(p+q-1),(p+q)}\big]$, and the random effect vector as $\bm{s}$. 
Then, Equation \eqref{Eq:1} can be written in matrix form as follows:
\begin{equation}\label{Eq:7}
    \bm{\eta} = \bm{X\beta} + \bm{Zu} + \bm{Wv} + \bm{s}.
\end{equation}

Assume that there may be $G$ subgroups within the data that explain potential sources of variability represented by $\bm{s}$ in Equation \eqref{Eq:7}, and index each subgroup by $g$ where $g=1,\dots,G$. 
Then, $\bm{s} = (\bm{s}_1,\ldots,\bm{s}_G)^T$ is the vector of random effects where $\bm{s}_{g}$'s are subgroup-specific random effects assumed to follow a particular distribution (typically Normal). 
If data are being collected across a spatial process, then Equation \eqref{Eq:7} could be extended to account for the fact that observations may not be independent. 
This could be achieved through an alternative specification of $\bm{s}$.  
For example, a latent spatial process could be assumed to describe underlying dependence within the data.  
Variability in the spatial process could then be described via a covariance function $C_{\kappa}(h)$ which defines the elements of a covariance matrix, where $h$ denotes the Euclidean distance between locations where data were collected and $\kappa$ denotes additional parameters, see  \textcite{diggle2007model}. 
The Matérn covariance function is commonly used throughout spatial modelling  \parencite{matern1966spatial}, and can be defined as follows:
\begin{equation}\label{Eq:8}
    C_{\kappa}(h)=\phi_1\{{2^{\kappa-1}\Gamma(\kappa)}\}^{-1}(h/\phi_2)^\kappa K_{\kappa}(h/\phi_2),
\end{equation}

\noindent where $K_{\kappa}(\cdot)$ is the modified Bessel function of order $\kappa$, $\phi_2>0$ is the scale parameter (determines the rate at which the correlation decays to zero with increasing $h$), $\kappa>0$ is a shape parameter (determines the analytic smoothness of the spatial process) and $\phi_1$ is the variance of the spatial process.


\subsection{Bayesian inference} \label{subsection:bayesianinference}

In this work, statistical inference will be conducted within a Bayesian framework.  
As such, all inferences are based on the posterior distribution of the model parameters $\bm{\theta}$. 
These unknown model parameters $\bm{\theta}$ are treated as random variables where {\it a priori} uncertainty about these parameters is represented by the probability distribution $p(\bm{\theta})$, which is known as the prior distribution. 
The prior distribution $p(\bm{\theta})$ and the likelihood $p(\bm{y}|\bm{\theta},\bm{x})$ of observing $\bm{y}$ given covariates $\bm{x}$ and parameter $\bm{\theta}$, can be combined using the Bayes' theorem as follows:
\begin{equation} \label{Eq:9}
    p(\bm{\theta}|\bm{y},\bm{x})=\frac{p(\bm{y}|\bm{\theta},\bm{x})p(\bm{\theta})}{p(\bm{y}|\bm{x})},
\end{equation}

\noindent where $p(\bm{\theta}|\bm{y},\bm{x})$ is the posterior distribution of $\bm{\theta}$ and $p(\bm{y}|\bm{x})$ is the normalising constant or the model evidence which ensures the posterior distribution integrates to one. 
Given the product of the above density/mass functions typically does not yield a known distribution, much of the research in Bayesian statistics has focused on developing efficient algorithms to sample from or approximate the posterior distribution. 
The most widely used approach for this purpose is Markov chain Monte Carlo (MCMC)  \parencite{metropolis1953equation}, with Hamiltonian Monte Carlo increasing in popularity  \parencite{girolami_calderhead_2011}.   

Often multiple models are considered for describing the observed data. 
Accordingly, model choice is a typical consideration in statistical inference. 
For this, assume that there are $J$ candidate models being contemplated for the data, and let $M$ denote the random variable associated with model $m$ such that $m=1,\ldots,J$. 
Each model $m$ is associated with model parameters $\bm{\theta}_m$, likelihood function $p(\bm{y}|M=m,\bm{\theta}_m,\bm{x})$, and prior distribution on parameters $\bm{\theta}_m$ denoted as $p(\bm{\theta}_m|M=m)$. 
Combining $p(\bm{y}|M=m,\bm{\theta}_m,\bm{x})$ and $p(\bm{\theta}_m|M=m)$ using the Bayes' theorem yields:
\begin{equation} \label{Eq:10}
    p(\bm{\theta}_m|M=m,\bm{y},\bm{x})=\frac{p(\bm{y}|M=m,\bm{\theta}_m,\bm{x})p(\bm{\theta}_m|M=m)}{p(\bm{y}|M=m,\bm{x})}.
\end{equation}

Formal Bayesian model choice requires the specification of prior model probabilities $p(M = m)$ and updating these based on observed data to form posterior model probabilities $p(M =m|\bm{y},\bm{x})$.  
Such probabilities can be evaluated as follows: 
\begin{equation}\label{Eq:11}
    p(M = m|\bm{y},\bm{x}) = \frac{p(M=m)p(\bm{y}|M =m,\bm{x})}{\sum_{j=1}^{J} p(M = j)p(\bm{y}|M = j,\bm{x})},
\end{equation}

\noindent where $p(\bm{y}|M=m,\bm{x})= \int_{\bm{\theta_m}} p(\bm{y}|M=m,\bm{\theta_m},\bm{x})p(\bm{\theta_m}|M=m)  \text{d} \bm{\theta}_m$ is the model evidence conditioned on model $m$, and there is a preference for the model with the largest posterior model probability. Analytical computation of the model evidence is generally only available for conjugate models in the exponential family. Thus, approximate methods are typically needed. For a review of available methods, see  \textcite{bos2002comparison}. 


\subsection{Bayesian inference for GA(M)Ms} \label{subsection:bayesianinferencegamm}

A Bayesian GA(M)M based on the O’Sullivan penalised splines  \parencite{wand2008semiparametric} and low-rank tensor product smoothers  \parencite{wood2016just} can be specified as follows:
\begin{equation}\label{Eq:12}
    \centering
    \begin{split}
     \bm{y} | \bm{\beta}, \bm{u}, \bm{v}, \bm{s} & \sim \mbox{EF}(\bm{\eta},\psi), \\ 
     g(\bm{\mu}) = \bm{\eta}  &= \bm{X\beta} + \bm{Zu} + \bm{Wv} +  \bm{s}, \\
     \bm{\beta} &= (\beta_0, \ldots,\beta_{p+q})^T, \\
     \bm{u} &= (\bm{u}_1, \ldots,\bm{u}_{p})^T, \\
     \bm{u}_j &= (u_{j1},\ldots,u_{jK_j}),\, j=1,\ldots,p \\
     \bm{u}_j|\sigma^2_{u_{j}} &\sim \text{N}(0,\sigma^2_{u_{j}}),\, j=1,\ldots,p\\
     \bm{v}_r|\lambda_{rf} ,\bm{S}_{rf} &\sim \text{MVN}\Big(\mathbf{0},\big(\sum_{f=0}^{2}  \lambda_{rf} \bm{S}_{rf}\big)^{-1}\Big), \,r=1,\ldots,R\\
     \bm\lambda &= (\bm\lambda_{1},\ldots,\bm\lambda_{R})^T , \bm\lambda_{r} = (\lambda_{r0},\lambda_{r1},\lambda_{r2}) , \\
    \end{split}
\end{equation}

\noindent where $\lambda$'s are  smoothing parameters and each $\bm{S}_{rf}, r=1,\dots,R$ and $f=0,1,2$, is a positive semi-definite marginal penalty matrices (see  \textcite{wood2006low} and  \textcite{wood2016just} for more information), where $
R={\comb{(p+q)}{2}}=(p+q)!/2!(p+q-2)!$ is the total number of possible two-way interaction terms. 
Further, $\bm{u}_j$ is the wiggliness parameter vector for $j^{th}$ spline term and $\bm{v}_r$ is the wiggliness parameter vector for $r^{th}$ interaction term.
The specification of $\bm{s}$ will depend on the exact model being fit.
Let $\bm\phi$ denote the parameters associated with the variability of $\bm{s}$, then $\bm{s}$ could be specified as $\bm{s}\sim p(\bm{s}|\bm\phi)$, if $\bm{s}$ accounts for variability between subgroups or based on the covariance function given in Equation \eqref{Eq:8}.  Finally, a GAM can be specified as a special case of the above model where the random effects $\bm{s}$ have been omitted.

For the above GA(M)M, the posterior distribution of $\bm{\theta}$ and $\bm{\alpha}$ is the distribution of interest where $\bm{\theta}=(\bm{\beta},\psi,\bm{\sigma}^{2}_{u},\bm\lambda,\bm\phi)$, ${\bm\alpha=(\bm{u},\bm{v},\bm{s})}$ and $\bm{\gamma} = (\bm{\sigma}^2_{u},\bm{\lambda},\bm{\phi})$. 
Accordingly, the posterior distribution can be defined as follows:
\begin{equation}\label{Eq:13}
    p(\bm{y}|\bm{\beta},\psi,\bm{\alpha},\bm{x})p(\bm{\alpha}|\bm{\gamma})p(\bm{\theta}),
\end{equation}

\noindent where $p(\bm{y}|\bm{\beta},\psi,\bm{\alpha},\bm{x})$ is the conditional likelihood of observing data $\bm{y}$ based on covariates $\bm{x}$ given $\bm{\beta},\psi$ and $\bm{\alpha}$. 
This simplifies straightforwardly to the posterior distribution for a GAM by dropping the terms associated with $\bm{s}$.  
In this paper, for the motivating design problem, we will adopt vague priors calibrated by prior predictive checks for fitting proposed GAMMs to historical data.  
The posterior distribution from this model will then form the prior information that will be used to construct Bayesian designs.


\section{Bayesian design} \label{section:bayesiandesign}

A design $\bm{d}$ can be broadly defined as the values of input variables that are specified for data collection.  
Most typically this will be the values of covariates $\bm{x}$ specified to collect data $\bm{y}$.  
Alternatively, within spatial settings, the design could be the actual locations at which data and covariate information will be collected.  
The aim of Bayesian design is to specify $\bm{x}$ to achieve a specified experimental goal.  
This could be, for example, to maximise the posterior precision of a given parameter or maximise the accuracy of predictions across a spatial region.  
This experimental goal is encapsulated within a utility function which is defined next.


\subsection{Utility functions} \label{subsection:utility}

Once the experimental goal has been specified, a function known as a utility function can be formulated to quantify how well this experimental goal would be addressed if data $\bm{y}$ were observed based on design $\bm{d}$.  
We denote such a utility function as $u(\bm{d},\bm{y},\bm{\theta},\bm{\alpha})$, which indicates that it may also depend on other variables such as model parameters.  
The goal is then to find the design that would maximise this utility function i.e.\ address the experimental goal as well as possible.  
However, as $\bm{y}$ is not known before the experiment, the expectation is taken with respect to this and other unknowns (e.g.\ $\bm{\theta}$ and $\bm{\alpha}$).  This forms an expected utility which can be defined as follows:
\begin{equation}\label{Eq:14}
    \begin{split}
        U(\bm{d})    =  {}& E_{\bm{y},\bm{\theta},\bm{\alpha}} [u(\bm{d},\bm{y},\bm{\theta},\bm\alpha)] \\
                =  {}& \int_{\bm{Y}} \int_{\bm\Theta} \int_{\bm{A}} u(\bm{d},\bm{y},\bm{\theta},\bm\alpha) p(\bm{y},\bm{\theta},\bm\alpha|\bm{d}) \text{d}\bm\alpha \text{d}\bm{\theta} \text{d}\bm{y} \\
                = {}& \int_{\bm{Y}} \int_{\bm\Theta} \int_{\bm{A}} u(\bm{d},\bm{y},\bm{\theta},\bm\alpha) p(\bm{y}|\bm{\theta},\bm\alpha,\bm{d}) p(\bm{\theta},\bm\alpha|\bm{d}) \text{d}\bm\alpha \text{d}\bm\theta \text{d}\bm{y}.\\
    \end{split}
\end{equation}

The goal of Bayesian design can then be stated as finding $\bm{d}^* = \arg\max_{\bm{d}\in\mathcal{D}}U(\bm{d})$.  
Unfortunately, in most cases, there is typically no closed-form solution to the above expectation, so an approximation is required. 
The most commonly used approximation is Monte Carlo integration defined as follows:
\begin{equation}\label{Eq:15}
    \hat{U}(\bm{d})\approx \frac{1}{L}\sum_{l=1}^L u(\bm{d},\bm{y}_l,\bm{\theta}_l,\bm{\alpha}_l),
\end{equation}

\noindent where $\big\{\bm{y}_l,\bm{\theta}_l,\bm{\alpha}_l\big\}_{l=1}^L$ is a sample generated from the joint distribution of $\bm{y}$, $\bm{\theta}$ and $\bm{\alpha}$.  
Thus, throughout this paper, a Bayesian design will be found by maximising the above approximation to the expected utility through the choice of the design $\bm{d}$.

In the examples that follow in the next section, we focus on parameter estimation as our experimental goal. 
Given this, an appropriate utility function is known as the the Kullback-Leibler divergence (KLD)  \parencite{kullback195110} which measures the distance between the prior and posterior distribution of the parameters. Such a utility function can be defined as follows:
\begin{equation}\label{Eq:16}
    U(\bm{d},\bm{y}) = \int_{\bm\Theta} \int_{\bm{A}} \log\Big(\frac{p(\bm{\theta},\bm{\alpha}|\bm{y},\bm{d})}{p(\bm{\theta},\bm{\alpha})}\Big)p(\bm{\theta},\bm{\alpha}|\bm{y},\bm{d})\text{d}\bm\alpha\text{d}\bm\theta.
\end{equation}

From Equation \eqref{Eq:15}, it can be seen that approximating the expected utility requires sampling from or approximating a large number of posterior distributions. 
This renders the use of algorithms like MCMC computationally infeasible to use in Bayesian design for realistically sized problems. 
Thus, fast methods for approximating the posterior distribution are needed. For this, we propose to use the Laplace approximation, and this is described in the next section. 


\subsection{Fast approximation to the posterior distribution} \label{subsection:fastapprox}

As a computationally efficient approach to approximate the posterior distribution, we propose to use the Laplace approximation which has the following form:
\begin{equation*}\label{Eq:17}
    \bm{\theta}|\bm{y},\bm{d} \sim \text{MVN}\big(\bm{\theta}^{\ast},\bm{B}(\bm{\theta}^{\ast})^{-1}\big),
\end{equation*}

\noindent where $\bm{\theta}^*$ and the Hessian matrix $B(\bm{\theta}^{\ast})$ at $\bm{\theta}^*$ are defined as:
\begin{align}\label{Eq:18}
    \begin{split}
        \bm{\theta}^* = \argmaxA_{\bm\theta} \space \{\log p(\bm{y}|\bm\theta,\bm{d}) + \log p(\bm\theta)\} \,  \text{and}\\ 
        B(\bm{\theta}^{\ast}) = {\frac{-\partial^2 \{ \log p(\bm{y}|\bm\theta,\bm{d}) + \log p\bm\theta)  \}}{ \partial \bm\theta \partial \bm{\theta}'}} \Bigl\lvert_{\bm{\theta}=\bm{\theta}^*}.
    \end{split}
\end{align} 

The above approximation requires evaluating the full data likelihood (i.e.\ not the conditional likelihood).  Finding this likelihood requires integrating out the  the wiggliness and random effects as follows: 
\begin{equation}\label{Eq:19}
    \begin{split}
       p(\bm{y}|\bm\theta,\bm{d}) = \int_{A} p(\bm{y}|\bm\beta,\psi,\bm{\alpha},\bm{d})p(\bm{\alpha}|\bm\gamma)  \text{d}\bm\alpha.
    \end{split}
\end{equation}

To obtain this for a GAM, we need to integrate out the $u$'s and $v$'s as follows:
\begin{equation}\label{Eq:20}
    p(\bm{y}|\bm\theta,\bm{d})= \int_{\bm{V}}\int_{\bm{U}} \prod_{i=1}^{n} p(y_i|\bm\beta,\psi,\bm{u},\bm{v},\bm{d}_i) p(\bm{u}|\bm{\sigma}^2_u) p(\bm{v}|\bm\lambda) \text{d}\bm{u} \text{d}\bm{v},
\end{equation}

\noindent where $p(y_i|\bm\beta,\psi,\bm{u},\bm{v},\bm{d}_i)$ is the conditional likelihood of observing $y_i$ at design point $\bm{d}_i$. 
Similarly, for a GAMM, if $\bm{s}$ is group specific random effects, we need to integrate out $u$'s, $v$'s and $s$'s as follows:
\begin{equation}\label{Eq:21}
    p(\bm{y}|\bm{\theta},\bm{d})=\int_{\bm{S}} \int_{\bm{V}}\int_{\bm{U}} \prod_{i=1}^{n} \prod_{g=1}^G p(y_{ig}|\bm\beta,\psi,\bm{u},\bm{v},s_{g},\bm{d}_i) p(\bm{u}|\bm{\sigma}^2_u) p(\bm{v}|\bm{\lambda}) p(\bm{s}|\bm\phi) \text{d}\bm{u} \text{d}\bm{v} \text{d}\bm{s},
\end{equation}

\noindent where $p(y_{ig}|\bm\beta,\psi,\bm{u},\bm{v},s_{g},\bm{d}_i)$ is the conditional likelihood of observing $y_{ig}$ at design point $\bm{d}_i$.

Once we have $\bm{\theta}^{\ast}$, it can be used to find a computationally efficient approximation to the posterior of the wiggliness and random effect parameters, $\bm{\alpha}$. 
This exploits the conditional independence between the model parameters and random effect terms, and thus just requires approximating the marginal posterior distribution of the wiggliness and random effect parameters (as the marginal posterior distribution of the model parameters is given by the above Laplace approximation). 
Let's denote the random variable associated with marginal posterior of $\bm{\alpha}$ given $\bm{\theta}^\ast$ by $\bm{\alpha}_{\theta}^\ast$. Then the marginal posterior can be found as follows:
\begin{equation*}\label{Eq:23}
    \bm{\alpha}_{\theta^\ast} | \bm{y},\bm{d} \sim \text{MVN}\big(\bm{\alpha}_{\bm{\theta}^{\ast}}^{\ast},H(\bm{\alpha}_{\bm{\theta}^{\ast}}^{\ast})^{-1}\big),
\end{equation*}

\noindent where $\bm{\alpha}_{\bm{\theta}^{\ast}}^{\ast}$ and the Hessian matrix $H(\bm{\alpha}_{\bm{\theta}^{\ast}}^{\ast})$ at $\bm{\alpha}_{\bm{\theta}^{\ast}}^{\ast}$ are defined as:
\begin{equation}\label{Eq:24}
\centering
    \begin{split}
        \bm{\alpha}_{\bm{\theta}^{\ast}}^{\ast} =  \argmaxA_{\bm{\alpha}_{\bm{\theta}^{\ast}}} \space \{ \log{ p(\bm{y}|\bm{\beta}^{\ast},\psi^{\ast}, \bm{\alpha}_{\bm{\theta}^{\ast}},\bm{d})} + \log{p( \bm{\alpha}_{\bm{\theta}^{\ast}}|\bm{\gamma}^{\ast})}\} \, \text{and} \\
        H(\bm{\alpha}_{\bm{\theta}^{\ast}}^{\ast}) =  {\frac{-\partial^2 \{ \log{ p(\bm{y}|\bm{\beta}^{\ast},\psi^{\ast}, \bm{\alpha}_{\bm{\theta}^{\ast}},\bm{d})} + \log{p( \bm{\alpha}_{\bm{\theta}^{\ast}}|\bm{\gamma}^{\ast})}\} }{ \partial  \bm{\alpha}_{\theta^\ast} \partial \bm{\alpha}_{\theta^\ast}'}} \Bigl\lvert_{ \bm{\alpha}_{\theta^\ast}=\bm{\alpha}_{\bm{\theta}^{\ast}}^{\ast}}.
    \end{split}
\end{equation}

The approximation to the posterior variance-covariance matrix of $(\bm{\theta}^*,\bm{\alpha}_{\bm{\theta}^{\ast}}^{\ast})$ is then block diagonal where there are two blocks; one for the model parameters and the other for the wiggliness and random effects parameters.

As described in Section \ref{subsection:bayesianinference}, the main drawback of using the model evidence $p(\bm{y}|M=m,\bm{d})$ for model selection is that it can be difficult to evaluate analytically. 
Fortunately, the Laplace approximation can also be used to provide a computationally efficient approximation. 
This is achieved by approximating  $p(\bm{y},\bm{\theta}_m|M=m,\bm{d})$ to the second-order around $\bm{\theta}^*_m$ by applying a Taylor series expansion which results in the following approximation (see Appendix \ref{appendix:modelevidence} for the derivation):
\begin{align} \label{Eq:22}
    \log p(\bm{y}|M=m,\bm{d}) =  & \log p(\bm{y}|\bm{\theta}^*_m,M=m,\bm{d}) + \log p(\bm{\theta}^*_m|M=m,\bm{d}) \\ &  + {\frac{T}{2}} \log (2\pi) 
    - {\frac{1}{2}} \log |\bm{B}(\bm{\theta}^{\ast}_m)|, \notag
\end{align}

\noindent where $T$ is the number of parameters in the model, and we note that the values of $\log p(\bm{y}|\bm{\theta}^*_m,M=m,\bm{d}) + \log p(\bm{\theta}^*_m,M=m,\bm{d})$ and $\bm{B}(\bm{\theta}^{\ast}_m)$ are readily available from the Laplace approximation. 
Thus, model comparison can be performed efficiently based on Equation \eqref{Eq:22} with an approximation to the model evidence. 
Throughout the examples we consider in this paper, our model choice procedure will proceed by evaluating the posterior model probability for all possible model combinations of predictors and their two-way interactions (when both main effects are present) where each model will be considered equally likely {\it a priori}. 
The model that yields the largest posterior model probability will be selected as the preferred model, and subsequently used to inform design. 

Another benefit of adopting the Laplace approximation is that the posterior distribution is Multivariate Normal.  
Thus, if the prior distribution is also a Multivariate Normal, then the KLD utility can be evaluated analytically as follows:
\begin{equation}\label{Eq:25}
    U(\bm{d},\bm{y}) = 0.5\times\Bigg(\text{tr}\Big( \bm{\Omega}_0^{-1} \bm{\Omega}_1\Big) + \Big(\bm{\mu}_0-\bm{\mu}_1\Big)^{T} \bm{\Omega}_0^{-1} \Big(\bm{\mu}_0-\bm{\mu}_1\Big) -T - \log{\Bigg(\frac{\text{det}(\bm{\Omega}_0)}{\text{det}(\bm{\Omega}_1)}\Bigg)}\Bigg),
\end{equation}

\noindent where $\bm{\mu}_0=(\bm{\theta}_0,\bm{\alpha}_0)$, $\bm{\mu}_1=(\bm{\theta}^*,\bm{\alpha}_{\bm{\theta}^{\ast}}^{\ast})$, $\bm{\Omega}_0$ and $\bm{\Omega}_1$ are the prior mean vector, posterior mean vector, prior variance-covariance matrix, and posterior variance-covariance matrix, respectively.  
In cases where the prior is not Multivariate Normal, methods of  \textcite{overstall2018approach} can be adopted such that Equation \eqref{Eq:25} can still be applied.

To summarise, pseudo-code for our approach to approximate the expected utility is provided in Algorithm \ref{Alg:1}.

\begin{algorithm}[H] \label{Alg:1}
\SetAlgoLined
 Initialise the prior information p($\bm{\theta},\bm{\alpha}$) using $(\bm{\theta},\bm{\alpha}) \sim \text{MVN}\big((\bm{\theta}_0,\bm{\alpha}_0),\bm{\Omega}_0\big)$ and the design $\bm{d}$. \\
 \For{$l$ = $1$ to $L$}{
 Draw $\bm{\theta}_l$ and $\bm{\alpha}_l$ from prior $p(\bm{\theta},\bm{\alpha})$. \\
 Given $\bm{\theta}_l$ and $\bm{\alpha}_l$, simulate data $\bm{y}_l$ at design $\bm{d}$ from the assumed GA(M)M model outlined in Equation \eqref{Eq:12}. \\
 Approximate $\bm{\theta}_l^*$ and $B(\bm{\theta}_l^*)$ for the simulated data using Equation \eqref{Eq:18}. \\
 Given $\bm{\theta}_l^*$, approximate $(\bm{\alpha}_{\bm{\theta}^{\ast}})_l^{\ast}$ and Hessian matrix $H\big((\bm{\alpha}_{\bm{\theta}^{\ast}})_l^{\ast}\big)$ using Equation \eqref{Eq:24}. \\
 Set the joint posterior $p(\bm{\theta},\bm{\alpha}|\bm{y},\bm{d})$ using $\bm{\theta},\bm{\alpha}|\bm{y},\bm{d} \sim \text{MVN}\big((\bm{\theta}^*,{\bm{\alpha}_{\bm{\theta}^{\ast}}}^{\ast})_l,(\bm{\Omega}_1)_l\big)$, where 
 $$(\bm{\Omega}_1)_l =\begin{bmatrix} 
            B(\bm{\theta}_l^*)^{-1} & \bm{0} \\
            \bm{0} & H\big((\bm{\alpha}_{\bm{\theta}^{\ast}})_l^{\ast}\big)^{-1}
 \end{bmatrix}.$$ \\
 Evaluate approximation to KLD utility $U(\bm{d},\bm{y}_l)$ using in Equation \eqref{Eq:25}.\\}
 Approximate the expected utility $\hat{U}(\bm{d}) = \frac{1}{L} \sum_{l=1}^{L} U(\bm{d},\bm{y}_l)$. \\
 \caption{Approximating the expected utility function of a design}
\end{algorithm}

First, initialise the prior information for the GA(M)M. 
If historical data are available, then one can use the posterior of the fitted model to these historical data as the prior for the design (line 1). 
Based on this prior, model parameters are simulated (line 3), and a prior predictive sample is then generated (line 4). 
Based on these data, the posterior distribution of $\bm{\theta}_l$ needs to be found. 
To do so, one needs to evaluate the likelihood as in Equations \eqref{Eq:20} and \eqref{Eq:21}. 
Exact evaluation of these integrals is generally not possible, thus Monte Carlo methods can be used to yield an approximation as follows:
\begin{equation}\label{Eq:26}
    p(\bm{y}|\bm{\theta}, \bm{d}) \approx \frac{1}{E} \sum_{e=1}^{E} \prod_{i=1}^{n} p(y_i|\bm\beta,\psi,\bm{u}_e,\bm{v}_e,\bm{d}_i) \, \text{and},
\end{equation}

\begin{equation}\label{Eq:27}
    p(\bm{y}|\bm{\theta}, \bm{d}) \approx \frac{1}{E} \sum_{e=1}^{E} \prod_{i=1}^{n} \prod_{g=1}^G p(y_i|\bm\beta,\psi,\bm{u}_e,\bm{v}_e,s_{g_e},\bm{d}_i),
\end{equation}

\noindent where $s_{g_e} \sim p(\bm{s}|\bm\phi)$, $\bm{u}_e \sim  \text{N}(0,\bm{\sigma}^2_{u} )$, $\bm{v}_{r_e} \sim \text{MVN}\Big(\bm{0},\big(\sum_{f=0}^{2} \lambda_{r_{f}} \bm{S}_{r_{f}}\big)^{-1}\Big)$ and $E$ is sufficiently large. 

With this approximation to the likelihood, the Laplace approximation can be used to find the posterior distribution of $\bm{\theta}_l$ (line 5). 
By applying an additional Laplace approximation, we can find the posterior distribution of $(\bm{\alpha}_{\bm{\theta}^*})_l$ (line 6). 
Using the output from line 5 and 6, we can approximate the joint posterior distribution of $\bm{\theta}$ and $\bm{\alpha}$ (line 7) from which the utility function can be evaluated (line 8).  

\subsection{Optimisation algorithm} \label{subsection:optimisation}

Given we are now able to approximate the expected utility, an approach to find the design that maximises this approximation is needed.  
Despite adopting a computationally efficient approximation to the posterior distribution (i.e.\ the Laplace approximation), evaluating the approximation to the expected utility is still computationally demanding.  
Thus, we require an optimisation algorithm that does not require a large number of function evaluations and one that can handle noisy expected utility evaluations (from the Monte Carlo approximation). 
Further, throughout our examples, we consider a range of design variables that could be either discrete or continuous in nature, so our adopted optimisation algorithm must be able to handle either or a combination of such variables.  
Accordingly, we used the coordinate-exchange (CE) algorithm proposed by  \textcite{meyer1995coordinate} to search through a discrete design space. 
Such an algorithm begins with a random design (a random selection of design points) and is then optimised one design point at a time. 
To do so, a given design point is substituted for all possible values and the expected utility is evaluated for each. 
If any of the `new' design points yields a larger expected utility than the current design, then the design point with the highest utility is retained in the design (potentially with a certain probability).  
If not, then the original design point is retained.  
This process is repeated for all design points, and then the whole process is then repeated until there is no substantial improvement in the expected utility or after a fixed number of iterations. 
Such an algorithm is straightforward to implement for discrete design spaces.

To extend the CE algorithm to search across continuous design spaces, we consider the approximate coordinate exchange (ACE) algorithm  \parencite{overstall2017bayesian}.  The extension is the use of a Gaussian process (GP) to emulate and optimise the expected utility surfaces for the one-dimensional optimisations in the CE algorithm.  
This is efficient as only a relatively small number of expected utility evaluations are needed to fit the GP.  Otherwise, the approach of ACE is similar to CE.

To find optimal designs for the illustrative example in Section \ref{section:examples}, we use the ACE algorithm as the design variables are continuous. 
In the motivating example, a mixture of continuous and discrete design variables are considered. 
Therefore, the CE and ACE algorithms are used in combination to find optimal designs. 
We also fix the random numbers within the approximation to the expected utility such that the function is deterministic.  This is purely for convenience as the optimisation is more computationally efficient with a deterministic utility.


\section{Examples} \label{section:examples}

The proposed approach for finding robust Bayesian designs is demonstrated in this section through two examples. 
The first is an illustrative example where properties of designs for a linear additive model are derived while the second example finds robust designs for monitoring sub-merged shoals (see Section \ref{subsection:motivating_example}). 
In each example, we explore the robustness of our designs with respect to potential alternative models.  
Importantly, when applying our approach to find these robust designs, the alternative models are not explicitly defined, as is required in many approaches proposed to form model robust designs.  Instead, these models form a set of possible models that could be observed under the defined GA(M)M. 


\subsection{Illustrative example} \label{subsection:example1}

Consider finding Bayesian designs under the following linear additive model:
\begin{equation}\label{Eq:28}
    \centering
    \begin{split}
    \bm{y} \mid \bm{\beta}, \bm{u} & \sim  \text{N}( \bm{X\beta} + \bm{Zu}, \sigma^2_\varepsilon),\\
    \bm{\beta} & = (\beta_0,\beta_1)^T,\\
    \beta_j & \sim \text{N}(0,10^2 ), j=1,2, \\
    \bm{u} & = (\bm{u}_1)^T,\\
    \bm{u}_1 & = (u_{11},\ldots,u_{1K}) ,\\ 
    \bm{u}_1 | \sigma^2_{u} & \sim \text{N}(0,\sigma^2_{u} ) . \\
    \end{split}
\end{equation}

Let $\mathscr{X}$ be the design space of $\bm{x}$. We consider $\mathscr{X}\in[-1,1]$ for this example and $\bm{x}$ is normalised to [0,1] when fitting the GAM models. 
Here, we are interested in how a design found under the KLD utility function might vary depending on the priors for $\sigma_{u}$ and $\sigma_{\varepsilon}$ and the specification of $K$.

\begin{figure}[H]
    \centering
    \includegraphics[scale=0.8]{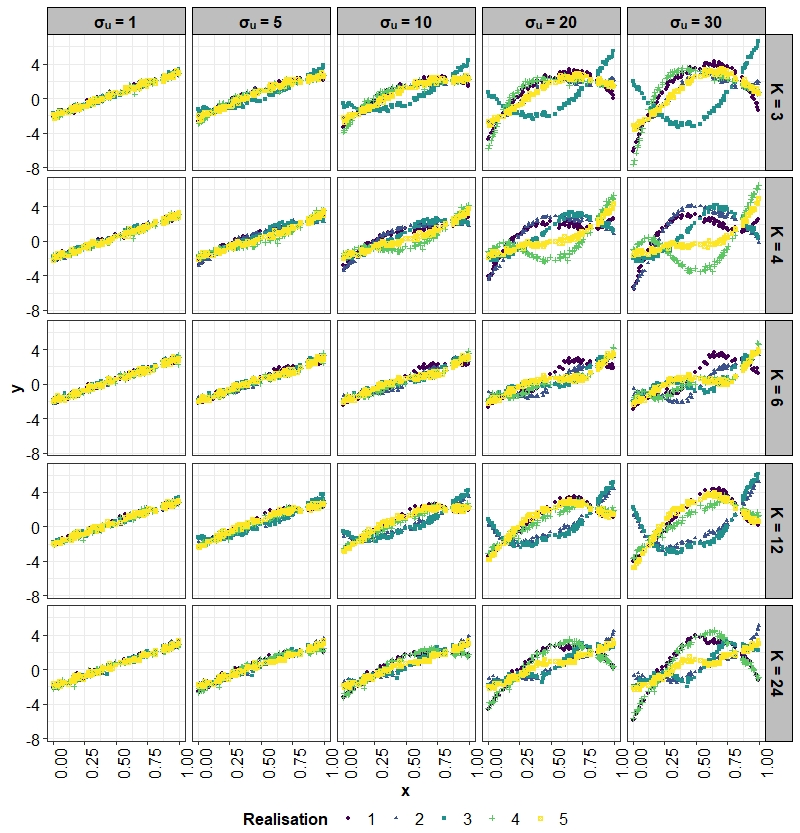}
	\caption{Five potential realisations that could be captured by GAM for same $\bm{\beta}$ values where each color represents a different realisation.}
    \label{figure:Ex1_wigg_data}
\end{figure}

To provide insight into the range of potential relationships between $\bm{x}$ and $\bm{y}$ that could be observed under the above linear additive model, realisations are shown in Figure \ref{figure:Ex1_wigg_data} for different values of $\sigma_{u}$ and $K$. 
For this, $\bm{y}$ were generated using the model in Equation \eqref{Eq:28} by randomly generating $\bm{x}$ and $\bm{u}$ while $\bm{\beta}$ was fixed at $\bm{\beta}=(2,-5)^T$. 

From Figure \ref{figure:Ex1_wigg_data}, when $\sigma_{u}$ is relatively low, only very minor deviations from the linear model are observed for all $K$.  
The flexibility of the model appears to come as this standard deviation increases, with moderate curvature apparent when $\sigma_{u} = 10$ and more extreme curvature observed when $\sigma_{u} = 20$ and $\sigma_{u} = 30$ . 
The flexibility of the model also appears to relate to $K$, with generally less flexibility (low wiggliness) observed when $K=3$ compared to larger values. 

To provide insight into the characteristics of Bayesian optimal designs under the KLD utility function under different parameter specifications of the model in Equation \eqref{Eq:28}, the following theorem has been derived for linear additive models (as defined above) with $p$ covariates (see Appendix \ref{appendix:prooftheorem1} for the proof).

\begin{theorem}\label{theorem_1}
Let $\bm{\theta}=\begin{bmatrix} \bm{\beta} & \bm{u}\end{bmatrix}$ be the vector of model parameters, $\bm{Q}=\begin{bmatrix}\bm{X} & \bm{Z}\end{bmatrix}$ be the design matrix, $\bm{\mu}_0$ be the prior mean vector, and $\bm{\Omega}_0=\begin{bmatrix} {\sigma^2_{\beta}} I_{p+1} & \bm{0} \\ \bm{0} & \blockdiagA_{1\le j\le k_p}\big({\sigma^2_{u_{p}}} I_{K_p}\big) \end{bmatrix}$ be the prior variance-covariance matrix for the linear additive model, then the joint posterior of $\bm{\beta}$ and $\bm{u}$ is 
\begin{equation*}
    \begin{bmatrix} \bm{\beta} & \bm{u} \end{bmatrix}^T \bigm| \bm{y}, \bm{Q} \sim \text{MVN}\big(\bm{\mu}_1, \bm{\Omega}_1 \big),
\end{equation*}

\noindent where $\bm{\mu}_1 = \big[\sigma^{-2}_{\varepsilon} \bm{Q}^T\bm{Q} + \bm{\Omega}_0^{-1}\big]^{-1} \big[\sigma^{-2}_{\varepsilon} \bm{Q}^T\bm{y}+ \bm{\Omega}_0^{-1}\bm{\mu}_0\big]$ and $\bm{\Omega}_1=\big[\sigma^{-2}_{\varepsilon} \bm{Q}^T\bm{Q} + \bm{\Omega}_0^{-1}\big]^{-1}$. 
\end{theorem}

Based on the result from Theorem \ref{theorem_1}, one can gain insight into the types of designs that would be preferred under different parameters configurations of the linear additive model described in Equation \eqref{Eq:28}. 
This was explored for a range of designs that varied in terms of how spread out the points are across the design space, see Appendix \ref{appendix:derivecoro1} for full details. 
The result from Theorem \ref{theorem_1} was then used to evaluate the expected KLD utility for these designs under different parameter configurations; the results of which are shown in Figure \ref{figure:theroy} of Appendix \ref{appendix:derivecoro1}.  
The general patterns in these results lead to the follow corollary.

\begin{coll}\label{Col:1}
Given the result from Theorem \ref{theorem_1}, the following patterns in preferred designs under the above linear additive model for the KLD utility function can be determined:  

\begin{enumerate}[label=(\roman*)]
  \item As $\sigma_{u}$ increases, there is a preference for more spread out design points.
  \item As $K$ increases, there is a preference for more spread out design points. 
  \item As $\sigma_{\varepsilon}$ increases, there is a preference for replicating design points.
\end{enumerate}
\end{coll}

The results of the above corollary would seem to make intuitive sense, particularly in light of the model realisations given in Figure \ref{figure:Ex1_wigg_data}.  
That is, as the wiggliness terms become more variable, design points become more spread out to estimate departures from the linear relationship.  
Indeed, when this term is relatively small, then a roughly linear relationship is observed so boundary points would be expected to be preferred.  
The same preferences would seem reasonable for increasing and decreasing values of $K$, respectively.  
Lastly, increasing replication as $\sigma_{\varepsilon}$ increases would also seem sensible for mean and variance estimation.  

\begin{figure}[H]
    \centering
    \includegraphics[scale=0.6]{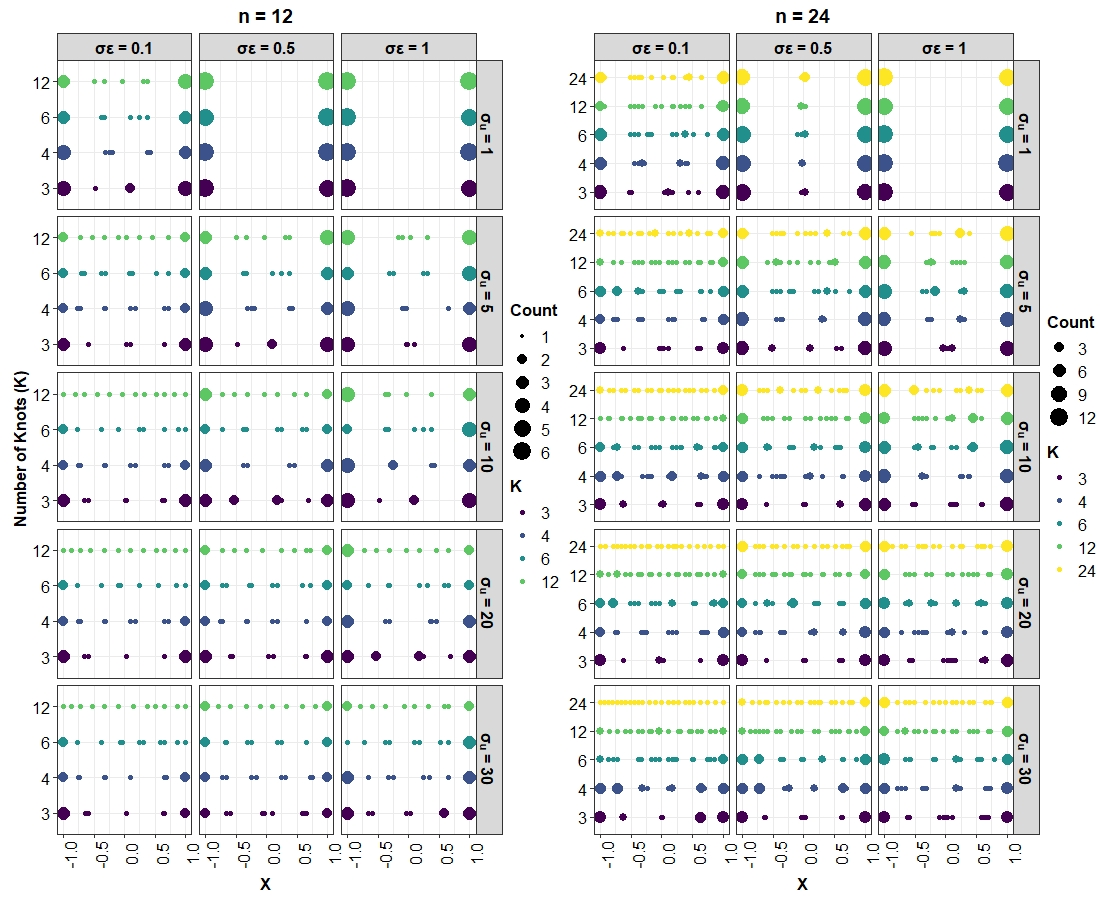}
	\caption{Optimal design points obtained for linear additive model under different values of $\sigma_{u}$, $K$, $\sigma_{\varepsilon}$, and $n$.}
    \label{figure:Ex1_design}
\end{figure}

To further explore designs under the KLD utility, we found optimal designs under a range of values for $\sigma_{u}$, $K$, $\sigma_{\varepsilon}$, and $n$ by applying the ACE algorithm described in Section \ref{subsection:optimisation}, and results are shown in Figure \ref{figure:Ex1_design}. 
Here, the values are set as $\sigma_{u}=(1,5,10,20,30)$, $K=(3,4,6,12)$ for $n=12$ and $K=(3,4,6,12,24)$ for $n=24$, and $\sigma_{\varepsilon}=(0.1,0.5,1)$. 
As can be seen, these results align with what is given in the above corollary, which is based on the result from Theorem \ref{theorem_1}. 

\begin{figure}[H]
    \centering
    \includegraphics[scale=0.55]{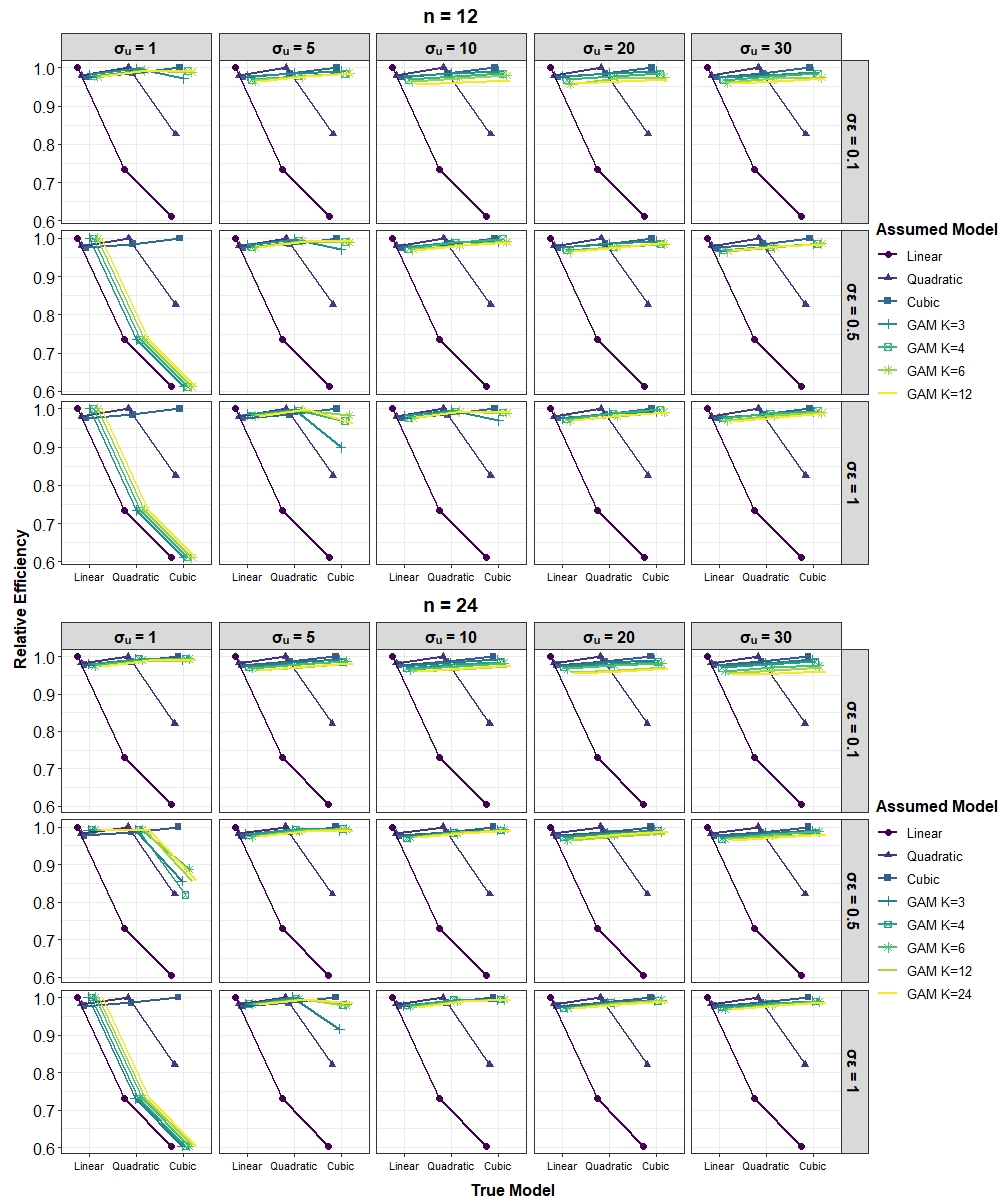}
	\caption{Relative efficiency ($R(\bm{d})$) of the optimal GAMM designs compared to optimal designs for polynomial models.}
	\label{figure:Ex1_rel_eff}
\end{figure}

To explore the robustness properties of the optimal designs shown in Figure \ref{figure:Ex1_design}, Bayesian designs under the KLD utility function were also found  under a linear, quadratic and cubic polynomial regression model. 
Here, an expression for the posterior was obtained following a similar approach as shown in the proof of Theorem 1, and the ACE algorithm was again used to find the optimal designs. 
Of note, such designs were found to be similar to those based on D-optimality, see  \textcite{atkinson2007optimum} and optimal designs were not influenced by the choice of $\sigma_\epsilon$.
Of interest is how well the designs found under the GAM would perform with respect to designs that would be optimal under these polynomial models.  
This is of interest as polynomial models are potential realisations under the GAM specification. 
To evaluate this, efficiency of a design $\bm{d}$ can be defined as:
\begin{equation}\label{Eq:29}
    R(\bm{d}) = \frac{U_{pol}(\bm{d})}{U_{pol}(\bm{d}_{pol}^*)},
\end{equation}

\noindent where $U_{pol}$ denotes the expected utility under a polynomial model and  $\bm{d}_{pol}^*$  is the optimal design under the polynomial model. 
If $\bm{d}$ is the optimal design under GAM model, $R(\bm{d})$ can be interpreted as the relative amount of information that is expected to be gained via a GAM design, compared to what would be optimal under the given polynomial model (if it is the preferred model).  
These efficiencies are shown in Figure \ref{figure:Ex1_rel_eff} for a range of different parameter configurations.  
As can be seen, $R(d)=1$ when one correctly guesses the appropriate underlying model, and this happens for the polynomial models as this is what was assumed in each of these cases. 
However, when this guess is incorrect, the loss in efficiency can be large, particularly when the assumed model is less complex than the underlying model.  
In terms of the designs found for the GAMs, all efficiencies remain above $0.9$ suggesting they are highly efficient for parameter estimation under each polynomial model.  
One exception is when $\sigma_{u}=1$. 
In this case, the behaviour of the model is close to that of a linear model (see Figure \ref{figure:Ex1_wigg_data}), thus the model is providing little robustness to departures from the linear relationship, resulting in reduced performance.

\subsection{Monitoring of submerged shoals} \label{subsection:example2}

As discussed in Section \ref{subsection:motivating_example}, our objective is to derive robust Bayesian designs for monitoring coral cover on submerged shoals. To form a basis for these designs, the initial surveys conducted by AIMS were considered.  
The response variable for this modelling,
coral cover, was assessed based on a series of images of the 
seafloor collected along a transect via towing an unmanned imaging rig approximately $1.5$ meters(m) off the seafloor. 
To convert an image into an estimate of coral cover, 20 points were randomly placed throughout each image, and then classified as being placed on coral or not.  This yielded Binomial data for each image, and led to the consideration of a Generalised additive logistic regression model.  
In terms of covariate information, an abundance of data about the reef topology were available.  
Full details are given in Appendix \ref{appendix:covariatedescription}. 
To account for potential spatial dependency in the data, a spatial grid was considered, see  \textcite{wines2020accounting} for further details. 
From this, the following GA(M)M can be proposed:
\begin{equation}\label{Eq:30}
    \centering
    \begin{split}
    \bm{y}\mid \bm{\beta}, \bm{u},\bm{v}, \bm{s} \sim {}& \text{Binomial}\Big(\text{logit}^{-1}\big(\bm{X}\bm{\beta} + \bm{Z}\bm{u} + \bm{W}\bm{v}+ \bm{s}\big),20\Big), \\ 
    \bm{\beta} &= (\beta_0, \ldots,\beta_{p+q})^T, \\
    \beta_j\mid \sigma^2_{\beta_j} & \sim \text{N}(0,\sigma^2_{\beta_j}), j=1,\ldots,p \\
     \bm{u} &= (u_1, \ldots,u_{p})^T, \\
     \bm{u}_j &= (u_{j1},\ldots,u_{jK_j}),\, j=1,\ldots,p \\
     \bm{u}_j|\sigma^2_{u_{j}} &\sim \text{N}(0,\sigma^2_{u_{j}}),\, j=1,\ldots,p\\
     \bm{v}_r| \lambda_{rf} ,\bm{S}_{rf} &\sim \text{MVN}\Big(\mathbf{0},\big(\sum_{f=0}^{2}  \lambda_{rf} \bm{S}_{rf}\big)^{-1}\Big), \,r=1,\ldots,R\\
     \bm\lambda &= (\bm\lambda_{1},\ldots,\bm\lambda_{R})^T , \bm\lambda_{r} = (\lambda_{r0},\lambda_{r1},\lambda_{r2}) , \\
     \bm{s} | \phi_1 & \sim {} \text{N}(0,\phi_1). \\
    \end{split}
\end{equation}

\begin{figure}[H]
        \centering
		\includegraphics[scale=0.5]{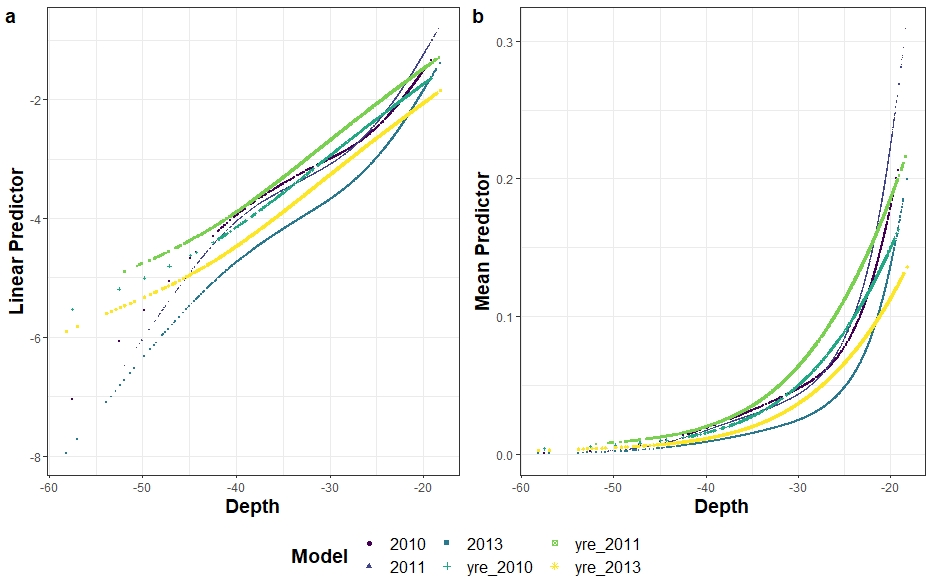}
		\caption{Relationship between the depth and predictor obtained using fitted GAMM through data from 2010, 2011, 2013 and all the data with year random effect (yre). Note that the fishnet random effects are not included in the predictions so that the relationship between depth and the linear predictor/mean prediction can be readily observed.}
        \label{figure:Ex2_depth_vs_predictor}
\end{figure}

As we have data from three different sampling years, we first considered data from each year separately. For each year, all possible combinations of covariates were considered along with all possible two-way interactions (if
the two main effects were included in the model). Note that, initially, before considering the models with two-way interactions, if
any pairwise correlation was greater than 0.5 (in absolute value), then that model was discarded. 
In addition, we also considered models for data collected across all years where a random effect (denoted as $\bm{t}$) was included to account for any inter-annual  variation.
This random effect $\bm{t}$ was included in the linear predictor in Equation \eqref{Eq:30} as an additive term, where we denote the variance of $\bm{t}$ by $\phi_2$.
To determine which covariates and two-way interactions were appropriate to include in this model, we followed the same procedure as described for the models for each year. 
Model choice was undertaken via the model evidence (Equation \eqref{Eq:22}), where each model was considered equally likely {\it a priori}. 
This resulted in the most appropriate model, for all cases, being the one that included depth as the only covariate. 
Depth is commonly a strong predictor of ecological patterns in marine environments  \parencite{barnes1999introduction}.  For corals, this relationship is largely due to changes in ambient light with depth  \parencite{laverick2020generalized}, as the corals considered here are all species that have zooxanthellae, and are therefore, at least in part, reliant on light for photosynthesis  \parencite{falkowski1984light}.
The relationship between depth and the mean prediction based on these four models is displayed in Figure \ref{figure:Ex2_depth_vs_predictor}. Based on each of these model fits, the posterior distribution from each model was considered as prior information for constructing sampling designs (see Appendix \ref{appendix:B_priors_for_design}). 

\begin{figure}[H]
    \centering
    \includegraphics[scale=0.5]{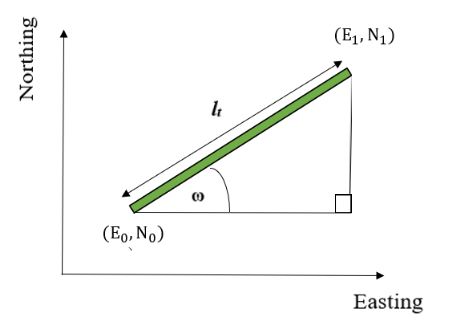}
    \caption{Orientation of a transect based on the design parameters; starting point ($E_0,N_0$), length $l_t$ and angle $\omega$.}
    \label{figure:Ex2_tr_ori}
    \vspace{-0.5cm}
\end{figure}

To find the Bayesian designs for each of these four preferred models, a search must be conducted across the whole shoal. 
To achieve this, we use a combination of CE and ACE algorithms where new sampling locations (i.e.\ different to those previously sampled) are possible. 
To search across the shoal, design parameters are introduced in order to define the placement of a transect. 
This is achieved by defining three design parameters; the starting point of the transect using two coordinates based on the Easting and Northing, say $E_0$ and $N_0$, the angle of the transect, say $\omega$ and the length of the transect, say $l_t$ in meters (m) (see Figure \ref{figure:Ex2_tr_ori}). 
 
Depending on the length of the transect, the end point (i.e.\ $E_1$ and $N_1$) can be evaluated using the following equations:
\begin{equation}\label{Eq:31}
\begin{split}
    E_1 = {}& E_0 + l_{t} \text{cos}(\omega),\\
    N_1 = {}& N_0 + l_{t} \text{sin}(\omega).\\    
\end{split}
\end{equation}

Based on the transects in the previously collected data and additional practical constraints, the total number, length, width and number of data points collected within each transect was set to 18, $500$m, $50$m and $50$, respectively.  
Discrete points were considered for $E_0$ and $N_0$ by introducing a $500\text{m} \times 500\text{m}$ spatial grid that covers the entire shoal and a continuous space was considered for $\omega$.  Thus, a CE algorithmic-type approach was used to search the space of $E_0$ and $N_0$ while an ACE algorithmic-type approach was used to search the space of $\omega$. 

\begin{figure}[H]
        \centering
		\includegraphics[scale=0.65]{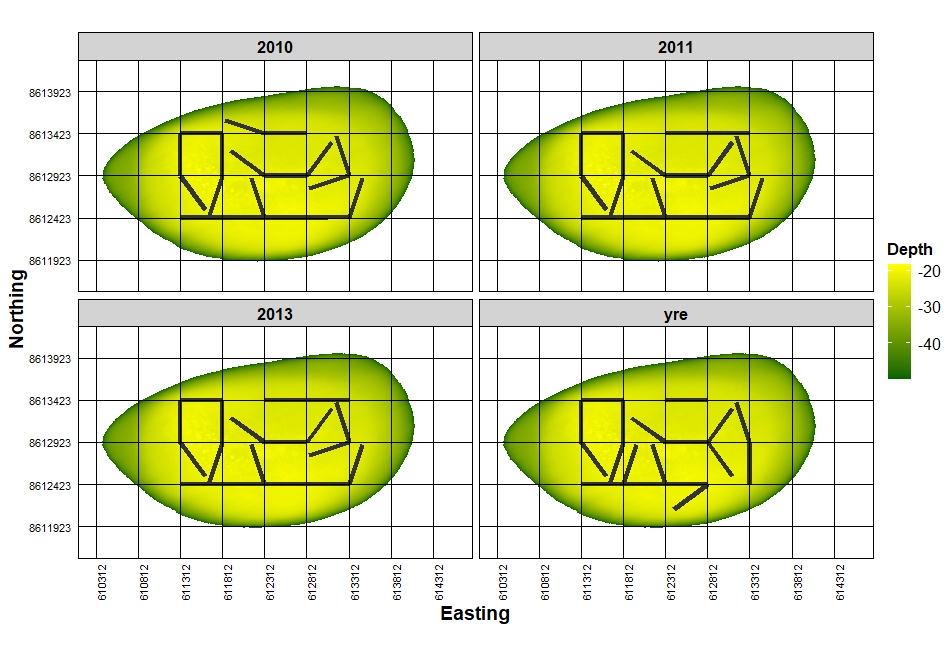}
		\caption{Optimal designs found under different priors displayed on the shoal. The $(500\text{m} \times 500\text{m})$ spatial grid which is used to define the design parameters is added to each plot.}
        \label{figure:Ex2_optimal_designs}
\end{figure}

Designs found under the four GAMMs are shown in Figure \ref{figure:Ex2_optimal_designs}. 
As can be seen, these designs appear to be located over relatively shallow areas of the shoal, and this is actually where the probability of having coral is high according to predictions from the fitted models (see Appendix \ref{appendix:C_maps}).  
In addition to this, data are also being collected across some gradients of depth, which would seem reasonable for estimating the associated effect.  

\begin{figure}[H]
        \centering
		\includegraphics[scale=0.65]{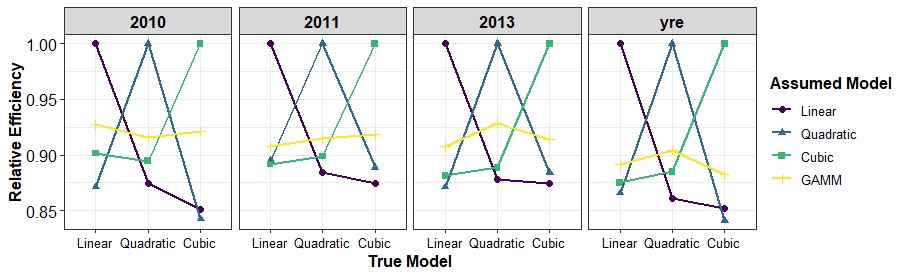}
		\caption{Relative efficiencies of the optimal GAMM designs compared to optimal designs obtained under polynomial models. Each plot represents a different prior used to find the optimal design based on data collected in different years while the "yre" means that there is a random effect component included in the model based on data collected across all three years.}
        \label{figure:Ex2_relative_efficiency}
\end{figure}

As in the previous example, we explored the robustness properties of the designs shown in Figure \ref{figure:Ex2_optimal_designs}.  
To do so, we again found Bayesian designs under alternative polynomial models of degrees one, two and three. Based on the relationship between depth and the logit of the mean coral cover found from the previously fitted models, these polynomial models do not seem unreasonable (see Figure \ref{figure:Ex2_depth_vs_predictor}). 
Thus, these polynomial models were each fit to the data sets with depth as a covariate (resulting in a total of $12$ models). 
Then we used the posterior from these models to form priors for the respective polynomial designs. 
Then, by applying the same optimisation algorithm used for the GAMM designs, we found the corresponding $12$ optimal designs. 
Accordingly, design efficiencies were calculated using Equation \eqref{Eq:29}, and are shown in Figure \ref{figure:Ex2_relative_efficiency}. 

From Figure \ref{figure:Ex2_relative_efficiency}, it can be seen that the designs found based on the GAMMs remain highly efficient (i.e.\ relative efficiency $>0.875$) under the three alternative models, and this can be a reasonable improvement when compared to assuming a polynomial model. This highlights the robustness properties of designs found under the approach proposed in this paper. 


\section{Discussion} \label{section:discussion}

Optimal sampling strategies are critical for deeper coral reef and shoal systems, due to not only the ecological importance as biodiversity hotspots but also the significant cost of accessing and field sampling in remote offshore and deeper water environments. To address this problem, we propose a  Bayesian design framework for finding designs that are robust to unknown model uncertainty.  The key innovation is rather than finding designs under a linear or generalised linear model, designs are found under a flexible model.  The flexible model considered here is a GAM, which is well-known for providing flexible model fits when the functional relationship between the response and continuous predictors is unknown.  Here, we have exploited this flexibility to provide model robust designs.  The benefits of doing so have been demonstrated in two examples where designs found under a GAM formulation were shown to be robust across different alternative models.  This is a highly desirable property of these designs as, of note, such alternative models were not explicitly specified {\it a priori}. Such robustness would seem appealing in practice, and this point may be supported by the recent surge in the use of GAMs for inference more generally.

Here our focus was on deriving designs for monitoring coral cover and demonstrating how GAM(M)s can be used to form robust designs when the underlying model is unknown. 
However, the approach developed could be used to explore a range of questions in the context of optimising a monitoring design for these off-shore submerged shoal habitats. 
Key questions of future interest in the study of these benthic communities include: 1) is there an optimal transect length and/or density of sampling points (images) within each transect that should be used; 2) what is the spatial scaling of patchiness of hard coral within shoal habitat, and should alternative modelling approaches be considered; 3) how does spatial precision impact optimal design configurations; and 4) what would be the optimal design for detecting changes through time. As interest in these submerged reef habitats grows  \parencite{kahng2010community}, developing optimal designs for detecting changes in their rich coral fauna through time is critical, which requires a coral cover per taxa approach, and extensions to our methods could be considered to address this. 

Across the examples considered in this work, we adopted the KLD utility function as our focus was on parameter estimation broadly across all model parameters, including parameterising the relationship with depth with greater certainty. However, there would also be value in focusing on developing methods that optimise alternative utilities. An example of interest would be maximising the precision of the estimated probability of observing coral across the shallow reef area of a shoal within each year. Such a utility may yield designs that are better suited to assessing the impact of potential catastrophic events, such as oil spills, coral bleaching or severe storm events.
Some insight into this can be provided from the designs found here but a utility could be constructed to target this explicitly.  
In principle, the approach proposed here could be considered for this purpose, and this is an area of research we plan to consider into the future.

While a large number of covariates were available for developing our model for the case study, model selection resulted in the clear outcome that across all years, and with year was included as a temporal random effect, a model containing only depth was appropriate. Depth represents a strong gradient in a range of environmental factors that can influence marine communities. In particular, the attenuation of light with depth is a key driver determining the distribution of organisms that rely on photosynthesis, such as the zooxanthellate corals  \parencite{laverick2020generalized}. As we would expect, the probability of observing coral was always highest on the shallowest parts of the shoal reaching values as high as 25 percent (see Figure \ref{figure:Ex2_depth_vs_predictor}). Such probabilities of observing hard coral are comparable to the more commonly recognised shallow reef communities. The prevalence of these zooxanthellate corals drops off rapidly with depth, declining to less than one percent predicted mean probability at depths of between 35 and 40 meters. The strong relationship with depth observed here and elsewhere suggests that a similar depth-based model framework may have wide applicability for determining optimal sampling designs for similar shoal-like mesophotic reefs more broadly. Future work could aim to explore the degree of model transferability  \parencite{yates2018outstanding} from the one developed for Barracouta shoal, to other shoals across the north-west shelf and elsewhere. However, it must be noted that the expected depth range of zooxanthellate corals will differ among regions due to variation in light attenuation because of differences in the optical properties of the water, such as turbidity. For example, in a comparison of the benthic communities at Glomar Shoal and Rankin Bank, also off the north-west coast of Australia, there was a difference of up to 20m in the lower limits of phototrophic taxa  \parencite{abdul2018biodiversity}. Given the strong dependence of phototrophic communities on light, future work may do better to focus on a light-driven process model, perhaps building on the recent work of  \textcite{laverick2020generalized}.

The focus of the current work has been on hard coral cover, which is a common metric used to assess health of coral reef communities  \parencite{obura2019coral}. A focus on hard coral cover only, is however, one limitation of this work, because it does not consider changes biodiversity nor community composition. New community assemblages with altered species composition may confer different functional traits. Designs suitable for monitoring community assemblages should be a priority for future modelling, to allow us to monitor a range of taxa of interest and/or conservation concern.

To provide additional robustness to model uncertainty, one could consider placing a prior distribution on $K$.  This was not pursued in this work as Example 1 was exploratory, and the choice for Example 2 was relatively clear.  However, we note that this would be a straightforward extension to include in future studies, as appropriate.  In addition, alternative flexible modelling approaches could be adopted (rather than a GAM implementation).  For example, a Gaussian Process model could be included within the linear predictor to capture model discrepancy  \parencite{kennedy2001bayesian}.  
It would seem that including such a term could lead to similarly model robust designs, and is an area we also hope to explore into the future.

\printbibliography[title=References,heading=bibnumbered]

\section*{Acknowledgments}
We thank the Seascape Health and Resilience team of the Australian Institute of Marine Science for the curation and provision of data.
We acknowledge the Aboriginal and Torres Strait Islander People as the Traditional Owners of the places where AIMS works, both on the land and in the sea country of tropical Australia. 
We pay our respects to the Elders; past, present and future; and their continuing culture, beliefs and spiritual relationships and connection to the land and sea.

\section*{Funding}
DDS is supported by the Australian Technology Network of Universities Industry Doctoral Training Centre (ATN IDTC) Scholarship with partner funding from the Australian Institute of Marine Science.
JMM was supported by an Australian Research Council Discovery Project (DP200101263).

\vspace{-0.6cm}

\begin{appendices}

\section{Additional Material for Section \ref{subsection:example1}}
\subsection{Proof of Theorem \ref{theorem_1}}\label{appendix:prooftheorem1} 

\begin{proof}

Let $\bm{y}= \bm{X}\bm{\beta} +\bm{Z}\bm{u} + \bm{\varepsilon}$,  $\bm{\beta}=\begin{bmatrix} \beta_0 & \beta_1 & \ldots & \beta_p \end{bmatrix}^T$,  
$\bm{u}=\begin{bmatrix}u_{11} & \ldots & u_{1{K_1}} & \ldots & u_{p1} & \ldots & u_{p{K_p}} \end{bmatrix}^T$, $\bm{y}=\begin{bmatrix} y_1 & \ldots & y_n \end{bmatrix}^T$, 
$\bm{X}=\begin{bmatrix} 1 & x_{11} \ldots & x_{1p}\\ \vdots & \vdots & \vdots \\ 1 &  x_{n1} \ldots & x_{np} \end{bmatrix}$, 
$\bm\varepsilon = \begin{bmatrix} \varepsilon_1 & \ldots & \varepsilon_n \end{bmatrix}^T$ and
$ \bm{Z} = \begin{bmatrix}
z_{11}(x_{11}) & \ldots & z_{1K_1}(x_{11}) & \ldots & z_{p_1}(x_{p_1}) & \ldots & z_{pK_p}(x_{p_1}) \\
\vdots & \ldots & \vdots & \ldots & \vdots & \ldots & \vdots \\
z_{11}(x_{1n}) & \ldots & z_{1K_1}(x_{1n}) & \ldots & z_{p_1}(x_{pn}) & \ldots & z_{pK_p}(x_{pn})
\end{bmatrix}$. Then the Bayesian model for the linear additive model can be written as follows:

\begin{align*}
\bm{y}\bigm| \bm\beta,\bm{u},\bm{X},\bm{Z},\sigma^2_{\varepsilon} &  \sim \text{N} \big( \bm{X}\bm{\beta}+\bm{Z}u,\sigma^2_{\varepsilon}\textbf{1}_n\big),  \\ 
\beta_j|\sigma^2_{\beta_j} & \sim \text{N}(0,\sigma^2_{\beta_j} ), j=1,\ldots,p \\ 
 \bm{u} &= (u_1, \ldots,u_{p})^T, \\ 
 \bm{u}_j &= (u_{j1},\ldots,u_{jK_j}),\, j=1,\ldots,p \\ 
 \bm{u}_j|\sigma^2_{u_{j}} &\sim \text{N}(0,\sigma^2_{u_{j}}),\, j=1,\ldots,p.\\ 
\end{align*}

\noindent To obtain the posterior of the model parameters, let $\bm{\theta}=\begin{bmatrix} \bm{\beta} & \bm{u}\end{bmatrix}$ be the vector of model parameters, $\bm{Q}=\begin{bmatrix}\bm{X} & \bm{Z}\end{bmatrix}$ be the design matrix, $\bm{\mu}_0$ be the prior mean vector, and  $\bm{\Omega_0}=\begin{bmatrix} {\sigma^2_{\beta}} I_{p+1} & \bm{0} \\ \bm{0} & \blockdiagA_{1\le j\le k_p}({\sigma^2_{u_{p}}} I_{K_p}) \end{bmatrix}$ be the prior variance-covariance matrix, then

\begin{align*}
  p\big(\bm\beta,\bm{u},\sigma^2_{\varepsilon},\sigma^2_{u},\sigma^2_{\beta} \bigm| \bm{y}, \bm{Q} \big) \ \propto \ & p\big(\bm{y}\bigm| \bm\beta,\bm{u},\sigma^2_{\varepsilon},\sigma^2_{u},\sigma^2_{\beta}, \bm{Q} \big)\ . \ p\big( \bm\beta,\bm{u},\sigma^2_{\varepsilon},\sigma^2_{u},\sigma^2_{\beta} \big)  \\
  \propto \ & p\big(\bm{y}\bigm| \bm\beta,\bm{u},\sigma^2_{\varepsilon}, \bm{Q} \big)\ . \ p\big( \bm\beta,\bm{u}, \bigm| \sigma^2_{\varepsilon}, \sigma^2_{u},\sigma^2_{\beta} \big) \ . \ p\big(\sigma^2_{\varepsilon},\sigma^2_{u},\sigma^2_{\beta} \big)  \\
  \propto \ & p\big(\bm{y}\bigm| \bm\beta,\bm{u},\sigma^2_{\varepsilon}, \bm{Q} \big)\ . \ p\big( \bm\beta,\bm{u}, \bigm| \sigma^2_{u}, \sigma^2_{\beta} \big) \ . \ p\big(\sigma^2_{\varepsilon},\sigma^2_{u},\sigma^2_{\beta} \big). 
\end{align*}

\noindent If $\sigma^2_{\varepsilon},\sigma^2_{u},\sigma^2_{\beta}$ are known,
\begin{align}\label{Eq:1_dash}
  p & \big( \bm\beta,\bm{u} \bigm| \sigma^2_{\varepsilon}, \sigma^2_{u}, \sigma^2_{\beta}, \bm{y}, \bm{Q} \big) \notag \\
 \propto \ & p\big(\bm{y}\bigm| \bm\beta,\bm{u}, \sigma^2_{\varepsilon}, \bm{Q} \big) .  p\big(\bm{\beta}, \bm{u} \bigm| \sigma^2_{u},\sigma^2_{\beta} \big) \notag \\
  \propto \ & \exp{\Bigg( \frac{(\bm{y}-\bm{Q}\bm{\theta})^T (\bm{y}-\bm{Q}\bm{\theta})}{-2\sigma^2_{\varepsilon}} \Bigg)} . \exp{\Bigg( \frac{(\bm{\theta} -\bm{\mu}_0)^T \bm{\Omega}_0^{-1} (\bm{\theta} -\bm{\mu}_0) }{-2} \Bigg)} \ 
  \{\because \ \bm{\theta}=\begin{bmatrix} \beta & u\end{bmatrix} \} \notag \\
  = \ & \exp{\Bigg(\frac{\sigma^2_{\varepsilon}(\bm{y}^T\bm{y} - \bm{y}^T\bm{Q}\bm{\theta} - \bm{\theta}^T\bm{Q}^T\bm{y} + \bm{\theta}^T\bm{Q}^T\bm{Q}\bm{\theta}) + \bm{\theta}^T\bm{\Omega}^{-1}_0\bm{\theta} - 2\bm{\theta}^T\bm{\Omega}^{-1}_0 \bm{\mu}_0 + \mu^T_0 \bm{\Omega}^{-1}_0 \bm{\mu}_0 }{-2}   \Bigg)} \notag \\
  = \ & \exp{\Bigg( -\frac{1}{2} \Big( \bm{\theta}^T \big(\sigma^{-2}_{\varepsilon}\bm{Q}^T\bm{Q} + \bm{\Omega}_0^{-1}\big)\bm{\theta} - 2\sigma^{-2}_{\varepsilon} \bm{\theta}^T \bm{Q}^T\bm{y}- 2\bm{\theta}^T\bm{\Omega}^{-1}_0 \bm{\mu}_0 \Big) \Bigg)} +  \mbox{constant}.
\end{align}
 
\noindent Assume that,
\begin{align}
\bm\beta,\bm{u} \bigm| \sigma^2_{\varepsilon},\sigma^2_{u},\sigma^2_{\beta}, \bm{y}, \bm{Q} \ \sim \ \mbox{MVN}(\bm{\mu}_1,\bm{\Omega}_1), \notag
\end{align}
where $\bm{\mu}_1$ is the posterior mean vector, and $\bm{\Omega}_1$ is the posterior variance-covariance matrix for the above linear additive model, then

\begin{align}\label{Eq:2_dash}
 p\big( \bm{\beta}, \bm{u} \bigm| \sigma^2_{\varepsilon},\sigma^2_{u},\sigma^2_{\beta}, \bm{y}, \bm{Q} \big) \propto &  \exp{\Big(-\frac{1}{2} (\bm{\theta}-\bm{\mu}_1)^T \bm{\Omega}_1^{-1} (\bm{\theta}-\bm{\mu}_1) \Big)} \notag \\
 = & \exp{\Big(-\frac{1}{2} (\bm{\theta}^T\bm{\Omega}^{-1}_1\bm{\theta} -2\bm{\theta}^T\bm{\Omega}^{-1}_1\bm{\mu}_1 + \mu^T_1\bm{\Omega}^{-1}_1\bm{\mu}_1)}\Big) \notag \\
 = & \exp{\Big(-\frac{1}{2} (\bm{\theta}^T\bm{\Omega}^{-1}_1\bm{\theta} -2\bm{\theta}^T\bm{\Omega}^{-1}_1\bm{\mu}_1)}\Big) + \mbox{constant}.
\end{align}

\noindent By comparing Equations \eqref{Eq:1_dash} and \eqref{Eq:2_dash}, 
\begin{align*}
\bm{\mu}_1 = & \ \big[\sigma^{-2}_{\varepsilon} \bm{Q}^T\bm{Q} + \bm{\Omega}^{-1}_0\big]^{-1} \big[\sigma^{-2}_{\varepsilon} \bm{Q}^T\bm{y}+ \bm{\Omega}^{-1}_0\bm{\mu}_0\big],  \\ 
\bm{\Omega}_1 = & \ \big[\sigma^{-2}_{\varepsilon} \bm{Q}^T\bm{Q} + \bm{\Omega}^{-1}_0\big]^{-1}. 
\end{align*}

\end{proof}

\clearpage
\subsection{Derivation of Corollary \ref{Col:1}} \label{appendix:derivecoro1} 
 
To derive the points $i-iii$ in Corollary 1, consider the numerical setup in following two tables. 
Table \ref{Tab:1} and \ref{Tab:2} form five different designs with $n=12$ and six different designs with $n=24$, respectively. 
For instance, in Table \ref{Tab:1}, design with index $1$ has $2$ unique design points $(-1,1)$, each repeated six times such that total number of design points are $12$. 
Further, all the designs are equally spaced and the distance between two consecutive points are displayed in the third column. 
The final column displays the unique design points of each design and the number of repetitions of each unique design point.
 
 \begin{table}[H]
    \small
    \caption{Design setup for Corollary 1 when $n=12$}
    \label{Tab:1}
    \begin{tabular}{cccc} \hline
    \textbf{Design Index} & \textbf{$\mbox{Points} \times \mbox{Repetitions}$} & \textbf{Design (n=12)} & \textbf{Distance} \\ \hline
    1 & $2 \times 6$ & $(-1, 1) \times 6$ & $\frac{2}{1} = 2$ \\
    2 & $3 \times 4$ & $(-1, 0 ,1) \times$ 4 & $\frac{2}{2} =1$ \\
    3 & $4 \times 3$ & $(-1,-\frac{1}{3},\frac{1}{3},1) \times 3$ & $\frac{2}{3}$ \\
    4 & $6 \times 2$ & $(-1,-\frac{3}{5}, -\frac{1}{5}, \frac{1}{5}, \frac{3}{5},1) \times 2$ & $\frac{2}{5}$ \\
    5 & $12 \times 1$ & $(-1,-\frac{9}{11}, -\frac{7}{11}, -\frac{5}{11}, -\frac{3}{11},-\frac{1}{11},\frac{1}{11}, \frac{3}{11}, \frac{5}{11}, \frac{7}{11}, \frac{9}{11}, 1) \times 1$ & $\frac{2}{11}$ \\ \hline
    \end{tabular}
 \end{table}
 
 \begin{table}[H]
 \small
    \caption{Design setup for Corollary 1 when $n=24$}
    \label{Tab:2}
    \begin{tabular}{cccc} \hline
    \textbf{Design Index} & \textbf{$\mbox{Points} \times \mbox{Repetitions}$} & \textbf{Design (n=24)} & \textbf{Distance} \\ \hline
    1 & $2 \times 12$ & $(-1, 1) \times 12$ & $\frac{2}{1} = 2$ \\
    2 & $3 \times 8$ & $(-1, 0 ,1) \times$ 8 & $\frac{2}{2} =1$ \\
    3 & $4 \times 6$ & $(-1,-\frac{1}{3},\frac{1}{3},1) \times 6$ & $\frac{2}{3}$ \\
    4 & $6 \times 4$ & $(-1,-\frac{3}{5}, -\frac{1}{5}, \frac{1}{5}, \frac{3}{5},1) \times 4$ & $\frac{2}{5}$ \\
    5 & $12 \times 2$ & $(-1,-\frac{9}{11}, -\frac{7}{11}, -\frac{5}{11}, -\frac{3}{11},-\frac{1}{11},\frac{1}{11}, \frac{3}{11}, \frac{5}{11}, \frac{7}{11}, \frac{9}{11}, 1) \times 2$ & $\frac{2}{11}$ \\
    6 & $24 \times 1$ & \begin{tabular}{@{}c@{}} $(-1,-\frac{21}{23}, -\frac{19}{23}, -\frac{17}{23}, -\frac{15}{23},-\frac{13}{23},-\frac{11}{23},-\frac{9}{23},-\frac{7}{23},-\frac{5}{23},-\frac{3}{23},-\frac{1}{23},$ \\ $\frac{1}{23}, \frac{3}{23}, \frac{5}{23}, \frac{7}{23}, \frac{9}{23},\frac{11}{23},\frac{13}{23},\frac{15}{23},\frac{17}{23},\frac{19}{23},\frac{21}{23}, 1) \times 1$ \end{tabular} & $\frac{2}{23}$ \\\hline
    \end{tabular}
 \end{table}
 
\begin{figure}[H]
    \centering
    \includegraphics[scale=0.7]{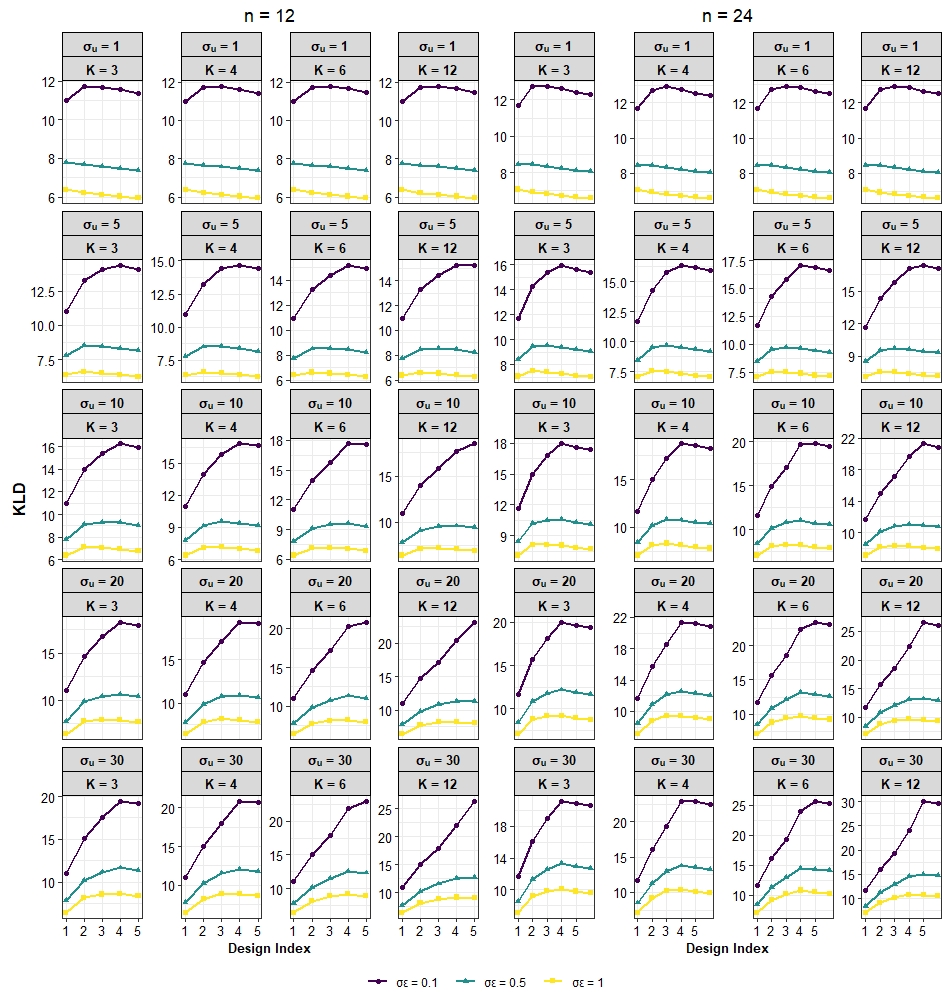}
	\caption{Left side panel displays KLD utility evaluations when $n=12$ (see Table \ref{Tab:1} for the design setup) while right side panel displays  KLD utility evaluations when $n=24$ (see Table \ref{Tab:2} for the design setup). }
    \label{figure:theroy}
\end{figure}

Then, we numerically evaluated the KLD utility  for the designs in Table \ref{Tab:1} and \ref{Tab:2} for different combinations of $K$ and $\sigma_u$ values (see Figure \ref{figure:theroy}). 
Here, KLD utility was evaluated based on the posterior we obtained in Theorem 4.1. 
These numerical results stand with points $i-iii$ of Corollary 1 for different combinations of $\sigma_u, K, \sigma_\varepsilon$ and $n$.

\section{Additional Materials for Section \ref{subsection:example2}}
\subsection{Priors for GAMM designs} \label{appendix:B_priors_for_design}

\begin{table}[H]
\center
\caption{Posterior mean and the standard deviation (s.d) of the model parameters from each GAMM model that were used as the prior for the design.}\label{Table:1}
    \begin{tabular}{cccc} \hline 
    \textbf{Prior} & \textbf{Parameters} & \textbf{Mean} & \textbf{s.d} \\  \hline 
    \multirow{4}{*}{2010} 
    & $\beta_0$ & $-6.66$ & $0.06$\\
    & $\beta_1$ & $\phantom{-}5.12$ & $0.08$ \\
    & $\log{(1/\sigma^2_{u}})$ & $-4.52$ & $0.01$\\
    & $\log{(1/\phi_1)}$ & $\phantom{-}3.40$ & $0.03$ \\  \hline 
    \multirow{4}{*}{2011}& $\beta_0$ & $-6.22$ & $0.01$ \\
    & $\beta_1$ & $\phantom{-}5.08$ & $0.01$\\
    & $\log{(1/\sigma^2_{u}})$ & $-3.40$ & $0.01$\\
    & $\log{(1/\phi_1)}$ & $\phantom{-}4.50$ & $0.01$ \\  \hline 
    \multirow{4}{*}{2013} & $\beta_0$ & $-7.77$ & $0.03$ \\
    & $\beta_1$ & $\phantom{-}6.04$ & $0.03$ \\
    & $\log{(1/\sigma^2_{u}})$ & $-3.77$ & $0.01$ \\
    & $\log{(1/\phi_1)}$ & $\phantom{-}4.14$ & $0.03$ \\  \hline 
    \multirow{4}{*}{yre} & $\beta_0$ & $-6.03$ & $0.005$ \\
    & $\beta_1$ & $\phantom{-}4.42$ & $0.01$\\
    & $\log{(1/\sigma^2_{u}})$ & $-2.49$ & $0.01$\\
    & $\log{(1/\phi_1)}$ & $\phantom{-}5.33$ & $0.02$ \\
    & $\log{(1/\phi_2)}$ & $\phantom{-}5.28$ & $0.02$ \\  \hline
    \end{tabular}
\end{table}


\subsection{Covariate description} \label{appendix:covariatedescription}

\begin{table}[H]
\caption{Description of covariates for the data used in monitoring shoals example.}
\centering
    \begin{tabular}{p{1.2in}p{1.2in}p{3.6in}} \hline 
    \multicolumn{1}{p{1.2in}}{\Centering { \textbf{Dataset prefixes used in spreadsheet}}} & \multicolumn{1}{p{1.2in}}{\Centering {\textbf{Predictor datasets}}} & \multicolumn{1}{p{3.6in}}{\Centering {\textbf{Definition}}} \\  \hhline{---} 
    \multicolumn{1}{p{1.2in}}{\Centering depth} & \multicolumn{1}{p{1.2in}}{\Centering Bathymetry} & \multicolumn{1}{p{3.6in}}{\Centering Elevation relative to the Australian Height Datum (AHD)} \\  
    \multicolumn{1}{p{1.2in}}{\Centering asp} & \multicolumn{1}{p{1.2in}}{\Centering Aspect} & \multicolumn{1}{p{3.6in}}{\Centering Azimuthal direction of the steepest slope, calculated on a $3 \times 3$ pixel area}  \\ 
    \multicolumn{1}{p{1.2in}}{\Centering slp} & \multicolumn{1}{p{1.2in}}{\Centering Slope} & \multicolumn{1}{p{3.6in}}{\Centering First\ derivative of elevation:  Average change in elevation / distance calculated on a $3 \times 3$ pixel area}  \\ 
    \multicolumn{1}{p{1.2in}}{\Centering prof} & \multicolumn{1}{p{1.2in}}{\Centering Profile curvature} & \multicolumn{1}{p{3.6in}}{\Centering Second\ derivative of elevation:  concavity/convexity parallel to the slope, calculated on a $3 \times 3$ pixel area}  \\ 
    \multicolumn{1}{p{1.2in}}{\Centering plan} & \multicolumn{1}{p{1.2in}}{\Centering Plan curvature} & \multicolumn{1}{p{3.6in}}{\Centering Second\ derivative of elevation:  concavity/convexity perpendicular to the slope, calculated on a $3 \times 3$ pixel area}  \\ 
    \multicolumn{1}{p{1.2in}}{\Centering curv} & \multicolumn{1}{p{1.2in}}{\Centering Curvature} & \multicolumn{1}{p{3.6in}}{\Centering Combined index of profile and plan curvature}  \\ 
    \multicolumn{1}{p{1.2in}}{\Centering hyp} & \multicolumn{1}{p{1.2in}}{\Centering Hypsometric index\textsuperscript{a}} & \multicolumn{1}{p{3.6in}}{\Centering Indicator of whether a cell is a high or low point within the local neighborhood}  \\ 
    \multicolumn{1}{p{1.2in}}{\Centering rng} & \multicolumn{1}{p{1.2in}}{\Centering Local relief (Range)\textsuperscript{ a}} & \multicolumn{1}{p{3.6in}}{\Centering Maximum minus the minimum elevation in a local neighborhood}  \\ 
    \multicolumn{1}{p{1.2in}}{\Centering std} & \multicolumn{1}{p{1.2in}}{\Centering Std Dev \textsuperscript{a}} & \multicolumn{1}{p{3.6in}}{\Centering Standard deviation of elevation}  \\ \hhline{---}
    \multicolumn{3}{p{\dimexpr6.06in+4\tabcolsep\relax}}{\flushleft \textsuperscript{a}{\fontsize{9pt}{10.8pt}\selectfont  Local neighborhood\ analysis:  run on circles of kernal pixel radius $5, 10,15,20,25,30,35,40,45,50$ original cell size is $6$m interpolated multibeam}}  \\ [-0.5em]
    \\ \hhline{---} \\ [-0.5em]
    \end{tabular}
\end{table}

\subsection{Maps of predicted probability} \label{appendix:C_maps}

\vspace{-0.6cm}
\begin{figure}[H]
    \centering
    \includegraphics[scale=0.7  ]{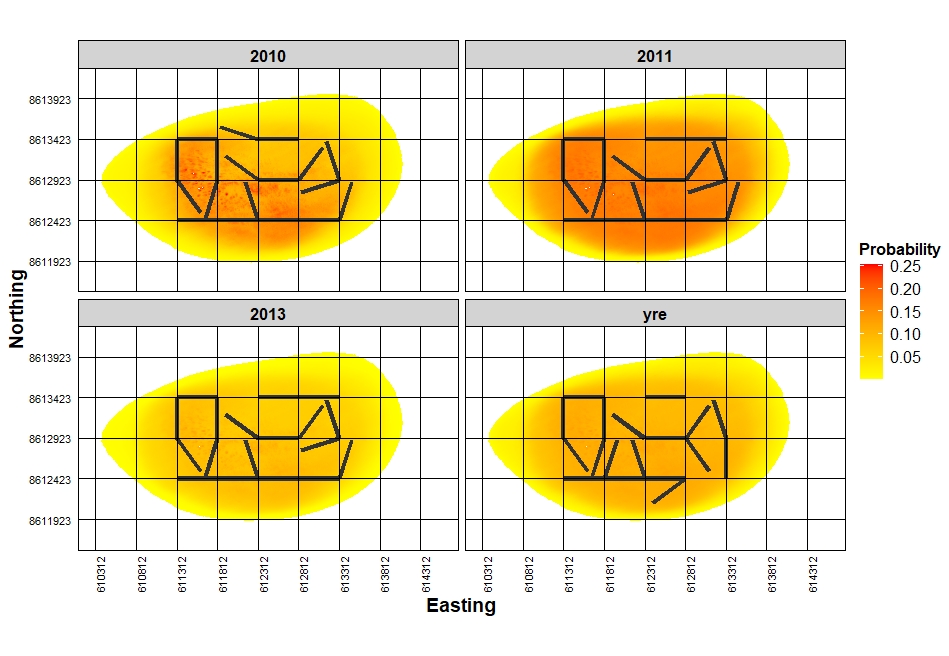}
	\caption{Maps of predicted probability of having coral based on the four fitted models which were used as the priors for the four designs.}
\end{figure}

\section{Derivation of Equation \ref{Eq:22}} \label{appendix:modelevidence}
\vspace{-0.6cm}
\begin{align*}
    p & (\bm{y} | M=m,\bm{d})\\
    = &  \int_{\bm{\theta}_m} p(\bm{y},\bm{\theta}_m | M=m,\bm{d}) \mbox{d}{\bm{\theta}_m} \notag \\
    = & \int_{\bm{\theta}_m}  \exp{\Big(\log{\Big(p(\bm{y},\bm{\theta}_m | M=m,\bm{d})\Big)}\Big)} \mbox{d}{\bm{\theta}_m} \notag \\
    \approx  & \int_{\bm{\theta}_m} \exp{\Big(\nabla \log{\big(p(\bm{y},\bm{\theta}^*_m | M=m,\bm{d})\big)}(\bm{\theta}_m-\bm{\theta}^*_m) + \frac{1}{2} (\bm{\theta}_m-\bm{\theta}^*_m)^T \nabla^2 \log{\big(p(\bm{y},\bm{\theta}^*_m | M=m,\bm{d})\big)} (\bm{\theta}_m-\bm{\theta}^*_m)\Big)} \\
    & \hspace{30mm} p(\bm{y},\bm{\theta}^*_m | M=m,\bm{d}) \mbox{d}{\bm{\theta}_m}   \hspace{10mm} ( \mbox{obtained by applying the Taylor series expansion} ) \\
    = & \quad p(\bm{y},\bm{\theta}^*_m | M=m,\bm{d}) \int_{\bm{\theta}_m} \exp{\Big(0 + \frac{1}{2} (\bm{\theta}_m-\bm{\theta}^*_m)^T |\bm{B}(\bm{\theta}^{\ast}_{m})| (\bm{\theta}_m-\bm{\theta}^*_m)\Big)} \mbox{d}{\bm{\theta}_m} \notag\\ 
    & \hspace{45mm} ( \because |\bm{B}(\bm{\theta}^{\ast}_{m})|=\nabla^2 \log{\big(p(\bm{y},\bm{\theta}^*_m | M=m,\bm{d})\big)}, \quad \bm{B}(\bm{\theta}^{\ast}_{m}) \quad\mbox{is the Hessian matrix}) \\
    = & \quad p(\bm{y},\bm{\theta}^*_m | M=m,\bm{d}) (2\pi)^{\frac{T}{2}}  |\bm{\Sigma}(\bm{\theta}^{\ast}_{m})|^{\frac{1}{2}} \int_{\bm{\theta}_m} \frac{1}{(2\pi)^{\frac{T}{2}} |\bm{\Sigma}(\bm{\theta}^{\ast}_{m})|^{\frac{1}{2}}} \exp{\Big(\frac{1}{2} (\bm{\theta}_m-\bm{\theta}^*_m)^T |\bm{\Sigma}(\bm{\theta}^{\ast}_{m})|^{-1} (\bm{\theta}_m-\bm{\theta}^*_m)\Big)} \mbox{d}{\bm{\theta}_m} \notag\\ 
    & \hspace{45mm} ( \because |\bm{B}(\bm{\theta}^{\ast}_{m})|=|\bm{\Sigma}(\bm{\theta}^{\ast}_{m})|^{-1} ) \\
    = & \quad  p(\bm{y},\bm{\theta}^*_m | M=m,\bm{d}) (2\pi)^{\frac{T}{2}} |\bm{B}(\bm{\theta}^{\ast}_{m})|^{-\frac{1}{2}} \times 1\notag \\
    = & \quad  p(\bm{y}| \bm{\theta}^*_m , M=m,\bm{d}) p(\theta^*_m| M=m,\bm{d}) (2\pi)^{\frac{T}{2}} |\bm{B}(\bm{\theta}^{\ast}_{m})|^{-\frac{1}{2}} \notag\\
\end{align*} 

$\therefore 
    \log p(\bm{y}|M=m,\bm{d}) = \log p(\bm{y}|\bm{\theta}^*_m,M=m,\bm{d}) + \log p(\bm{\theta}^*_m|M=m,\bm{d}) + {\frac{T}{2}} \log (2\pi) - {\frac{1}{2}} \log |\bm{B}(\bm{\theta}^{\ast}_{m})|.$
\end{appendices}

\end{document}